\documentclass[reprint,aps,prresearch,superscriptaddress]{revtex4-2}

\usepackage{dcolumn, makecell}

\usepackage{siunitx}

\usepackage{changes}
\usepackage{amsmath}
\usepackage{amssymb}
\usepackage{comment}

\usepackage{graphicx}% Include figure files
\usepackage{dcolumn}% Align table columns on decimal point
\usepackage{bm}% bold math
\usepackage{hyperref}
\hypersetup{colorlinks,allcolors=blue}
\usepackage{physics}
\usepackage{braket}
\usepackage{upgreek}
\usepackage{cleveref}

\usepackage[protrusion=true, expansion=true]{microtype} %better PDF typesetting
\newcommand{\ii}[0]{\textrm{i}}
\newcommand{\ee}[0]{\textrm{e}}

\newcolumntype{d}[1]{D{.}{.}{#1}}

\newcommand{\jump}[2]{\left[\!\left[#1\right]\!\right]_{#2}}

\begin{document}

%\preprint{APS/123-QED}

\title{Native multi-qubit gates on a single-junction unimon circuit}

\author{Sasu Tuohino}
    \affiliation{QCD Labs, Department of Applied Physics, Aalto University, P.O. Box 13500, FI-00076 Aalto, Finland}
\affiliation{Nano and Molecular Systems Research Unit, University of Oulu,\\ P.O.~Box 3000, FI-90014 Oulu, Finland}
    
% \altaffiliation{Physics Department, XYZ University.}%Lines break automatically or can be forced with \\
%\author{Vasilii Vadimov (?)}
%    \affiliation{QCD Labs, QTF Centre of Excellence, Department of Applied Physics, Aalto University, P.O. Box 15100, FI-00076 Aalto, Finland}

%\collaboration{MUSO Collaboration}%\noaffiliation

%\author{Wallace Teixeira (acknowledgement?)}
%    \affiliation{QCD Labs, QTF Centre of Excellence, Department of Applied Physics, Aalto University, P.O. Box 15100, FI-00076 Aalto, Finland}

%\collaboration{CLEO Collaboration}%\noaffiliation
%\author{Heikki Suominen (?)}
%    \affiliation{QCD Labs, QTF Centre of Excellence, Department of Applied Physics, Aalto University, P.O. Box 15100, FI-00076 Aalto, Finland}

\author{Mikko M\"ott\"onen}
    \affiliation{QCD Labs, Department of Applied Physics, Aalto University, P.O. Box 13500, FI-00076 Aalto, Finland}
    \affiliation{VTT Technical Research Centre of Finland Ltd., QTF Center of Excellence, P.O. Box 1000, FI-02044 VTT, Finland}
%\affiliation{Jointly supervised the work.}

\author{Matti Silveri}
\affiliation{Nano and Molecular Systems Research Unit, University of Oulu,\\ P.O.~Box 3000, FI-90014 Oulu, Finland}
%\affiliation{Jointly supervised the work.}
    
%\date{\today}% It is always \today, today,
             %  but any date may be explicitly specified

\begin{abstract}

Quantum processors with native multi-qubit gates may offer very efficient implementations of near-term quantum algorithms on noisy hardware. Here, we introduce the \textit{multiunimon}, a superconducting multimode circuit that encodes multiple qubits and enables native multi-qubit gates in a device consisting of a single Josephson junction embedded in a coplanar waveguide structure. Closely related to the unimon qubit, it inherits properties such as high anharmonicity, full protection against low-frequency charge noise, and partial protection against flux noise. By designing such a three-qubit device with Josephson-to-inductive energy ratio above unity and using a leakage-aware encoding scheme for the computational states, we simulate all twelve different controlled-controlled-NOT gates with a mean fidelity of 99.5\% with simple sine-squared pulses of comparable length to single-qubit gates. The performance is limited by incoherent errors dominated by dielectric loss. With improvements in noise protection, design, and pulse shaping, the simulations suggest that fidelities approaching 99.99\% are within reach. 
%Rather than following a predetermined labeling, the leakage-aware encoding selects the computational subspace directly from the device eigenspectrum to minimize gate errors. It also enables an extended gate set that enhances the connectivity between the computational states. 
Our results demonstrate the potential of the multiunimon as a highly connected multi-qubit unit for larger superconducting quantum processors.

\end{abstract}

%\keywords{Suggested keywords}%Use showkeys class option if keyword
                              %display desired
\maketitle

%\tableofcontents

\section{\label{sec:intro}Introduction}
%%%%%%%%%% Introduction %%%%%%%%%%

A practical quantum computer requires a universal set of quantum gates that can be executed with high fidelity~\cite{Nielsen_Chuang_2010}. For superconducting quantum processors~\cite{arute_quantum_2019, wu_strong_2021, gao_establishing_2025, acharya_quantum_2024}, universal control is typically achieved with single-qubit rotations driven by microwave pulses~\cite{krantz_quantum_2019}, combined with entangling two-qubit operations, activated either by frequency tuning---often via a tunable coupler~\cite{yan_tunable_2018, mundada_suppression_2019, stehlik_tunable_2021, sung_realization_2021, marxer_long-distance_2023, ding_high-fidelity_2023}---or by microwave drives on fixed-frequency qubits~\cite{rigetti_fully_2010, chow_simple_2011, sheldon_procedure_2016, paik_experimental_2016, mitchell_hardware-efficient_2021}. While these approaches work well, they tend to increase hardware complexity when scaled to larger systems. For instance, many tunable couplers require additional Josephson junctions and flux lines, adding a level of complexity comparable to that of another qubit~\cite{yan_tunable_2018}. Thus from the hardware point of view, multi-qubit devices, and the native gates they enable, appear as an attractive alternative.

Although the combination of single- and two-qubit gates is enough to realize any unitary operation, errors from imperfect gates and undesired coupling to the environment limit the circuit depth that quantum computers can reliably execute. This is especially true in the on-going era of noisy intermediate-scale quantum devices~\cite{preskill_quantum_2018}, where a diverse set of native gates can enable more efficient compilation of quantum algorithms and considerably improve the algorithmic performance. In particular, quantum error correction protocols~\cite{reed_realization_2012, rasmussen_single-step_2020, inada_measurement-free_2021, perlin_fault-tolerant_2023, heusen_measurement-free_2024} and many quantum algorithms, including Shor's~\cite{vedral_quantum_1996, haner_factoring_2017, gidney_how_2021, skosana_demonstration_2021} and Grover's~\cite{figgatt_complete_2017} algorithms, benefit from access to native multi-qubit gates. A canonical example of a native multi-qubit gate is the controlled-controlled-NOT (CCNOT) gate, i.e., the Toffoli gate. Although it is conceptually just a CNOT with an additional control qubit, its decomposition takes at least five two-qubit gates~\cite{yu_five_2013}. For instance, a decomposition using CNOT gates requires six CNOTs, two Hadamards, and seven T gates~\cite{shende_cnot-cost_2009}. Moreover, the CCNOT gate is a non-Clifford gate, and together with the Hadamard gate, it forms a universal gate set~\cite{aharonov_simple_2003,shi_both_2003}.

Native three-qubit gates have been realized in superconducting circuits through several distinct approaches, including multi-qubit coupler elements~\cite{mezzacapo_many-body_2014,glaser_controlled-controlled-phase_2023,simakov_high-fidelity_2024,liu_direct_2025}, simultaneous two-body interactions~\cite{gu_fast_2021,kim_high-fidelity_2022,baker_single_2022,nagele_effective_2022,warren_extensive_2023,itoko_three-qubit_2024}, and strongly interacting multimode circuits~\cite{roy_implementation_2017, roy_multimode_2018, roy_programmable_2020, maurya_universal_2026, pedersen_native_2019, christensen_scheme_2023, rasmussen_single-step_2020} such as the trimon circuit~\cite{roy_implementation_2017, roy_multimode_2018, roy_programmable_2020, maurya_universal_2026}. In a trimon, four Josephson junctions in a ring arrangement form three normal modes with strong always-on cross-Kerr interactions. This all-to-all coupling naturally enables a three-qubit device with native generalized controlled-controlled rotations CC$\mathcal{R}(\theta,\phi)$ driven by resonant microwave pulses. While the angle $\theta$ determining the population transfer is controlled by the pulse amplitude and length, the angle $\phi$ corresponding to controlled-controlled-Z (CCZ) rotations can be adjusted by the phase of the drive pulse. These CCZ rotations can be viewed as updates of the rotating frame tracked by the control software, also known as virtual Z gates, and thus are practically instant and error-free~\cite{mckay_efficient_2017}. This set of native gates is universal.
%\textcolor{orange}{For instance, controlled and single-qubit rotations can be realized with either simultaneous or sequential CC$\mathcal{R}$s~\cite{roy_multimode_2018}.}

The trimon is based on the transmon qubit~\cite{koch_charge-insensitive_2007}, which is the most used superconducting qubit variant today and has demonstrated impressive single-qubit~\cite{li_error_2023, marxer_above_2025} and two-qubit~\cite{marxer_above_2025, marxer_long-distance_2023,li_realization_2024} gate fidelities. However, the transmon has inherent weaknesses; most notable is its weak anharmonicity, which causes leakage to non-computational states and thereby limits both gate speed and single-qubit gate fidelity~\cite{krantz_quantum_2019, hyyppa_reducing_2024}. Despite its relatively low sensitivity to low-frequency charge noise, the sensitivity is not fully suppressed, leaving the qubit susceptible to parity switching events caused by quasiparticle tunneling~\cite{papic_charge-parity_2024}.

The unimon qubit was proposed and demonstrated to address these issues~\cite{hyyppa_unimon_2022}. The cancellation of the linear inductive terms increases the anharmonicity, and the absence of superconducting islands provides full protection against low-frequency charge noise. Further parameter optimization has shown that the unimon can be tuned to allow single-qubit gate fidelities above 99.99\%~\cite{duda_parameter_2025}. The unimon circuit consists of a Josephson junction embedded in a coplanar waveguide (CPW) resonator that is grounded at both ends, forming a gradiometric structure that suppresses symmetric flux noise while allowing the qubit frequency to be tuned by the differential flux between the two loops.

The CPW structure makes the unimon circuit inherently multi-modal, with important consequences for the unimon qubit in terms of renormalizing effects~\cite{tuohino_multimode_2024}. The linear CPW modes interact with the Josephson junction, producing shared nonlinear couplings between the modes that can be enhanced by placing the junction asymmetrically. With a well-chosen junction location, these Kerr-type couplings give the unimon a multi-mode structure resembling that of the trimon, but with only a single junction. The interaction strength, however, is primarily limited by the high frequencies of the CPW modes and by the requirement to keep the Josephson-to-inductive energy ratio below unity,
%$E_\text{J}/E_L \leq 1$, 
which avoids the double-well potential~\cite{hyyppa_unimon_2022} in which normal modes become ill-defined and difficult to treat numerically~\cite{tuohino_multimode_2024}.

In this work, we propose the \textit{multiunimon}, a single-junction superconducting circuit that natively supports twelve high-fidelity three-qubit gates with simulated $\pi$-rotation average fidelities of 99.56\% at \SI{20}{\nano\second} gate times. The fidelities are primarily limited by dielectric loss. The multiunimon is a single-junction analog of the transmon-based trimon, with a native gate set built from a similar family of CC$\mathcal{R}$ rotations. It extends the unimon circuit of Ref.~\cite{tuohino_multimode_2024} by adding a junction-free CPW line that intersects the original at a branching point, yielding a richer design space that allows lower frequencies for the CPW modes and stronger couplings between them.

The qubits in the multiunimon are defined by selecting eigenstates that minimize a metric estimating the infidelity of the native gate set, which is often dominated by leakage to non-computational states and off-resonant transitions within the computational subspace. We refer to this approach as \textit{leakage-aware encoding}. In contrast to the normal-mode encoding of trimons, the leakage-aware encoding does not follow a predetermined labeling based on dressed single-mode excitations. Instead, it is determined by a search through the Hilbert space for an optimal mapping from physical eigenstates to computational states that minimizes gate errors. This freedom is especially valuable in the strongly anharmonic regime, 
%$E_\text{J}/E_L \geq 1$, 
where normal-mode encoding breaks down. The combination of the encoding and the enhanced drive connectivity of this regime also yields an additional set of native gates, including a high-fidelity $\pi$ rotation between the computational states $\ket{000}$ and $\ket{111}$, implemented with an average fidelity of 99.74\% at the gate time of \SI{10}{\nano\second}. This gate enables single-step preparation of the Greenberger--Horne--Zeilinger (GHZ) state.

The design optimization is implemented with a simulation framework that searches over both, the circuit geometry and the qubit encoding, based on an iterative diagonalization scheme that provides fast and accurate evaluation of candidate designs. This framework can be used to explore the design space even further in search of better designs. In particular, designs with improved protection against dielectric loss could yield average three-qubit gate fidelities above 99.9\% at \SI{20}{\nano\second} gate times, approaching 99.99\% at \SI{60}{\nano\second}, as indicated by our gate simulations in the absence of decoherence.

The remainder of this work is structured as follows. In Section~\ref{sec:device}, we introduce the proposed circuit design and its Hamiltonian, and lay out the basic operations available for the multiunimon. Section~\ref{sec:encoding} explains the leakage-aware encoding and the connectivity benefits of operating in the strongly anharmonic regime with Josephson-to-inductive energy ratio exceeding unity.
%$E_\text{J}/E_L>1$.
Section~\ref{sec:optimization} covers the workflow used to optimize the design, and Section~\ref{sec:results} presents the simulated results for an example design. Finally, in Section~\ref{sec:summary_discussions}, we summarize the main results and discuss future research directions of the multiunimon not addressed here, such as readout and scaling.

\section{\label{sec:device}Multiunimon device}

\subsection{\label{sec:circuit_description}Circuit description}

\begin{figure*}[t]
\includegraphics[scale=0.9]{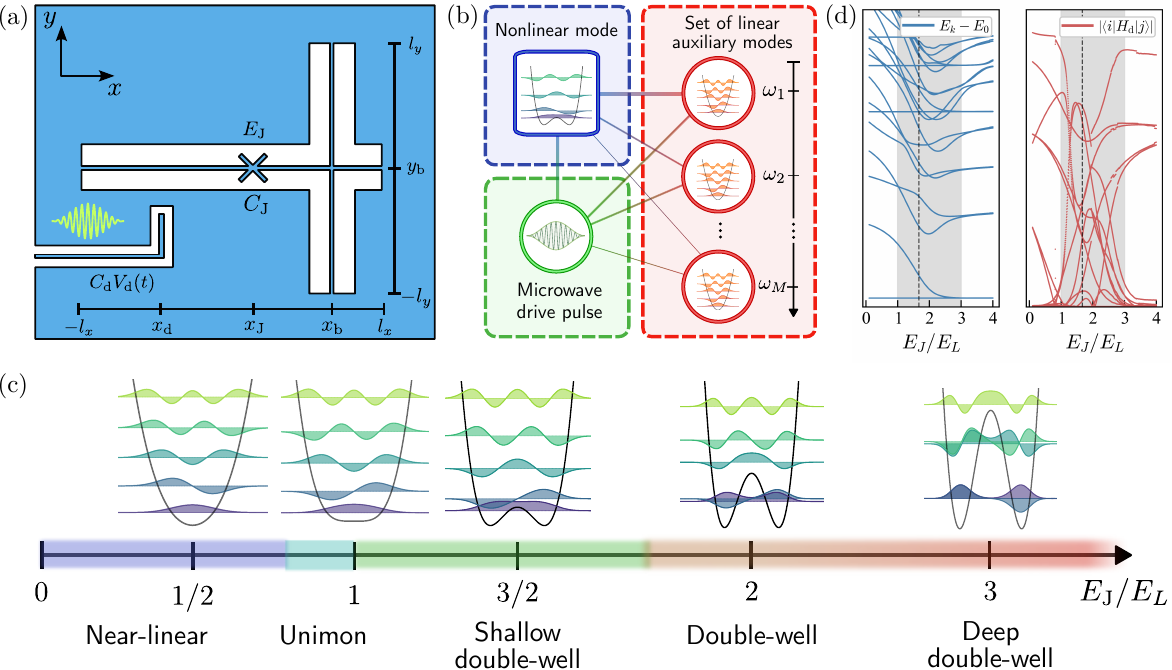}
\caption{\label{fig:circuit_schematic}
(a) Schematic top view of a single-junction multiunimon device with a capacitively coupled drive line. Blue color denotes thin-film superconducting metal and white the regions the absence of metal. The cross at $(x_\textrm{J}, y_\textrm{b})$ denotes the Josephson junction. (b) Bare mode structure of the auxiliary-mode model. The auxiliary modes couple to the nonlinear mode but not to each other, while the microwave drive couples to all modes. The coupling decreases for higher-frequency modes. (c) Potential landscape of the nonlinear mode in different regimes of the Josephson-to-inductive energy ratio $E_\text{J}/E_L$. (d) Eigenenergies and drive matrix elements as a function of $E_\text{J}/E_L$. The gray area highlights the regime of enhanced drive connectivity within the double-well region, before the barrier becomes too large and some eigenstates become nearly degenerate with transitions between them forbidden by symmetry. The dashed gray line indicates the $E_\text{J}/E_L$ value of the example design in Section~\ref{sec:results}.
}
\end{figure*}

The multiunimon is an extension of the original unimon circuit~\cite{hyyppa_unimon_2022, tuohino_multimode_2024}. It consists of a single Josephson junction embedded in a CPW structure at position $x_\text{J}$, with an additional CPW line of length $2l_y$ intersecting the main line of length $2l_x$ at the point $(x_\text{b},y_\text{b})$ as depicted in Fig.~\hyperref[fig:circuit_schematic]{\ref*{fig:circuit_schematic}(a)}. Each line is grounded at both ends, eliminating charge islands and protecting the circuit against low-frequency charge noise. The intersecting geometry forms four inductive loops, but the Josephson junction is involved in only two of them. As in the original unimon, the superconducting phase across the junction is therefore sensitive only to the difference of the external fluxes between the two adjacent loops.

% Maybe we should add the \Phi_\text{diff} to the figure and explain it in the text.

We divide the Hamiltonian of the multiunimon into four parts,
\begin{align}\label{eq:full_H}
    \hat H = \hat H_\text{JJ} + \hat H_\text{CPW} + \hat H_\text{int} + \hat H_\text{d}(t),
\end{align}
describing the nonlinear junction mode, the $M$ linear CPW modes, the interaction between the linear modes and the nonlinear mode, and a drive term that couples to every mode through geometry-dependent strengths [Fig.~\hyperref[fig:circuit_schematic]{\ref*{fig:circuit_schematic}(b)}], respectively.

The nonlinear mode is modeled as a Josephson junction with a capacitive and an inductive shunt, using Hamiltonian
\begin{align}\label{eq:H_JJ}
    \hat H_\text{JJ} = 4 E_C \hat n^2 + \frac{1}{2}E_L \hat \varphi^2 - E_\text{J} \cos (\varphi - \varphi_\text{diff}),
\end{align}
where $\hat n$ and $\hat \varphi$ are the conjugate charge and phase operators of the nonlinear mode, $E_\text{J}$ is the Josephson energy, and $E_C$ and $E_L$ are the effective charging and inductive energies, which are mainly determined by the CPW geometry. The charge and phase operators follow the canonical commutation relation $[\hat \varphi, \hat n]=\ii$. Throughout this work, we operate at the external-flux sweet spot $\varphi_\text{diff}=\pi$, where the contributions to the linear inductance partially cancel out~\cite{hyyppa_unimon_2022}.

The linear auxiliary modes of the CPW and their coupling to the junction take the form
\begin{align}
    &\hat H_\text{CPW} = \sum_{m=1}^M \hbar \omega_m \hat a^\dagger_m \hat a_m, \label{eq:H_CPW} \\  
    &\hat H_\text{int} = - \ii\hat n \sum_{m=1}^M \hbar g_m (\hat a_m - \hat a^\dagger_m),
\end{align}
where $\hat a_m$ and $\hat a_m^\dagger$ are the annihilation and creation operators of the mode $m$, $\omega_m$ is its angular frequency, and $g_m$ is the coupling strength to the nonlinear mode.

The coupling is capacitive, which is a consequence of the charge gauge~\cite{roth_optimal_2019, mehta_down-conversion_2023} used in the derivation of the Hamiltonian. This choice differs from the flux gauge of the auxiliary-mode model in Refs.~\cite{hyyppa_unimon_2022, tuohino_multimode_2024}, although the underlying physics is unchanged. The practical advantage of the charge gauge is that the nonlinear-mode parameters, mainly $E_L$, are less sensitive to the mode cutoff $M$, enabling more efficient numerical evaluation.

We implement the microwave drive by capacitively coupling the drive line to the CPW at a location $x_\text{d}$ chosen so that the drive couples to every desired transition in the computational subspace with sufficient strength. The drive Hamiltonian takes the form
\begin{align}
    &\hat H_\text{d}(t) = 2 e V_\text{d}(t) \biggl[\eta_0 \hat n + \ii\sum_{m=1}^M \eta_m \bigl( \hat a_m^\dagger - \hat a_m \bigr) \biggr],
\end{align}
where $V_\text{d}(t)$ is the time-dependent voltage on the drive line, with $\eta_0$ and $\eta_m$ denoting the geometry-dependent coupling strengths of the nonlinear and linear modes. 

\begin{figure*}[t]
\includegraphics[scale=0.9]{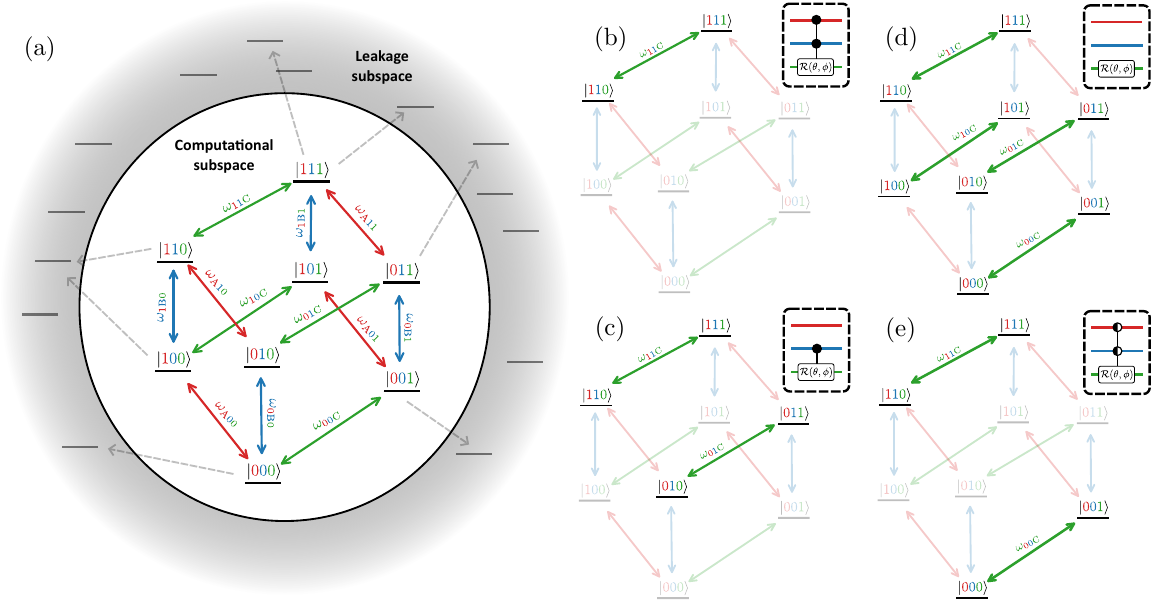}
\caption{\label{fig:cube_schematic}
(a) Three-qubit computational subspace of the multiunimon and the twelve native CC$\mathcal{R}$ gate connections within it. The computational states form a cube, with vertices corresponding to computational states and edges to transitions activated by native gates. Edges are color-coded by the target qubit; each target qubit has four edges, one per control-qubit configuration. Faint dashed arrows indicate possible leakage out of the computational subspace. (b) A microwave pulse at frequency $\omega_{11\text{C}}$ rotates the target qubit only if both control qubits are in state $\ket{1}$. This corresponds to a single CC$\mathcal{R}$ gate. (c) Adding a drive at frequency $\omega_{01\text{C}}$ removes the conditionality on one of the control qubits, producing a C$\mathcal{R}$ gate. (d) Driving all four frequencies for the same target qubit produces an unconditional single-qubit rotation. (e) Driving both frequencies $\omega_{11\text{C}}$ and $\omega_{00\text{C}}$, which correspond to control-qubit configurations of the same parity, produces a parity-controlled rotation, where the rotation is conditioned on the parity of the control qubits.
}
\end{figure*}

The branched geometry of the CPW does not change the form of the Hamiltonian, but it does affect its parameters, opening a richer design space. In particular, the additional flexibility in the mode frequencies $\omega_m$ and the coupling strengths $g_m$ enables high-fidelity multiunimon designs that are inaccessible to single-line geometries. We give the full derivation of the Hamiltonian in Appendix~\ref{app:Derivation}, where we also introduce a perturbative treatment for the elimination of the high-frequency modes. 

The mode structure of the Hamiltonian has a star topology in which the nonlinear mode acts as a central hub coupled to a set of auxiliary CPW modes, as illustrated in Fig.~\hyperref[fig:circuit_schematic]{\ref*{fig:circuit_schematic}(b)}. The coupling strengths $g_m$ decrease with increasing mode frequency, as also observed in Refs.~\cite{malekakhlagh_cutoff-free_2017, parra-rodriguez_quantum_2018}, suppressing the contributions of high-frequency modes. We exploit this structure in an iterative diagonalization scheme that diagonalizes the multiunimon Hamiltonian efficiently in regimes where direct exact diagonalization would be infeasible (Appendix~\ref{app:NRG}).

\subsection{\label{sec:basic_gates}Basic gate operations}

Inspired by the trimon, the multiunimon relies on a set of native microwave-driven CC$\mathcal{R}(\theta, \phi)$ gates, where the rotation angle $\theta$ is set by the drive amplitude and length, and the drive phase $\phi$ sets the rotation axis in the $xy$-plane of the target-qubit Bloch sphere. The gate set is general in the sense that for each target qubit there are four frequency-selective gates corresponding to each configuration of the control qubits. In total this adds up to twelve native three-qubit gates (Fig.~\ref{fig:cube_schematic}). By applying them simultaneously or sequentially, the conditionality can be reduced leading to C$\mathcal{R}(\theta,\phi)$ and $\mathcal{R}(\theta,\phi)$ gates for two and four  CC$\mathcal{R}$s, respectively [Fig.~\hyperref[fig:cube_schematic]{\ref*{fig:cube_schematic}(b)}--~\hyperref[fig:cube_schematic]{\ref*{fig:cube_schematic}(d)}]. Also gates such as parity-controlled rotations in Fig.~\hyperref[fig:cube_schematic]{\ref*{fig:cube_schematic}(e)}~\cite{christensen_scheme_2023, guo_parity-controlled_2025} and recently demonstrated Raman-based controlled-$\sqrt{\ii \text{SWAP}}$ and controlled-$\sqrt{\ii \text{bSWAP}}$ gates~\cite{maurya_universal_2026} are available by driving multiple pulses simultaneously. The $\ii \text{SWAP}$ and $\ii \text{bSWAP}$ denote the exchange between states within the odd- and even-parity two-qubit subspaces, respectively.

To make the gate set universal, controlled-controlled-phase (CCPHASE) rotations can be implemented efficiently via frame updates in the control software, which appear as adjusted drive phases $\phi$ for subsequent pulses~\cite{roy_multimode_2018, roy_programmable_2020, maurya_universal_2026}. These virtual Z gates~\cite{mckay_efficient_2017} cost no physical drive time and are limited only by the precision of the control hardware. Similar to CC$\mathcal{R}$s, they also yield the full hierarchy of conditional phase rotations, from a CCPHASE gate at the three-qubit level down to a single-qubit PHASE gate~\cite{roy_multimode_2018}.

The virtual Z gates are also used to remove undesired phase accumulations during the CC$\mathcal{R}$ gates. Since every rotation is implemented via a single drive line, the device is susceptible to off-resonant driving of spectator transitions in the computational subspace and of leakage transitions to non-computational states [Fig.~\hyperref[fig:cube_schematic]{\ref*{fig:cube_schematic}(b)}(a)]. Although the multiunimon is designed with sufficient frequency detuning from all undesired transitions to suppress leakage, ac Stark shifts induced by the drive cause phase accumulation in spectator states and over- or under-rotation on the target transition. 
%\textcolor{orange}{(Fig.~X)}
These errors are deterministic and can be reliably mitigated through careful calibration~\cite{maurya_universal_2026, tripathi_benchmarking_2025, wu_simultaneous_2025}: virtual Z gates correct the phase errors, while shaped pulses and adjusted drive frequencies correct rotation-angle errors. In this work, our simulations include virtual-Z corrections for spectator phase errors. For simplicity, we omit corrections for rotation-angle errors, which could yield modest improvements in the gate fidelities.

\section{\label{sec:encoding}Encoding for a high-fidelity gate set}

A competitive native three-qubit gate set for multiunimon requires eight computational eigenstates that are spectroscopically distinct. In Ref.~\cite{tuohino_multimode_2024}, we explored the physics of the single-line unimon circuit and observed Kerr-type couplings between its normal modes, but the couplings were weaker than those of the trimon and the single-line geometry could not support three low-frequency modes coupling simultaneously to the junction. The branched CPW geometry of the multiunimon resolves this frequency limitation, and the iterative diagonalization scheme grants access to the strongly anharmonic regime in which the circuit nonlinearity becomes sufficient. We now describe this regime and the encoding procedure that exploits it.

\subsection{Strongly anharmonic regime \texorpdfstring{$E_\text{J}/E_L >1$}{EJ/EL >1}}

The double-well regime, $E_\text{J}/E_L > 1$, has remained largely unexplored in unimon circuits because normal-mode-based numerical approaches break down there~\cite{hyyppa_unimon_2022}. The Hamiltonian of Eq.~\eqref{eq:full_H} does not rely on normal modes and is therefore well-suited for this parameter regime. By exploring it, we identify a sub-region with properties particularly useful for the multiunimon, distinct from both the unimon regime and the fluxonium-like double-well regime with a large barrier separating the wells [Fig.~\hyperref[fig:circuit_schematic]{\ref*{fig:circuit_schematic}(c)}]. This intermediate regime features a shallow double-well potential for the nonlinear mode, which increases the anharmonicity and enhances the charge matrix elements between states that are only weakly connected in the unimon regime [Fig.~\hyperref[fig:circuit_schematic]{\ref*{fig:circuit_schematic}(d)}].

In the unimon regime, the charge operator $\hat n$ acts approximately as a ladder operator on the low-energy spectrum, coupling each eigenstate primarily to its nearest neighbor in the energy ladder. Transitions involving multiple excitations, such as those described by $\mel{i+3}{\hat n}{i}$, are relatively weak. In the regime $E_\text{J}/E_L > 1$, the double-well potential produces dressed eigenstates with substantial weight across multiple bare energy levels, and these matrix elements involving multiple excitations become significant as we show below based on numerical evidence. Combined with the linear couplings to the CPW modes, the multiunimon thus supports a richer set of transitions accessible to the drive than the typical unimon circuit. However, the parity symmetry of the flux operating point $\varphi_\text{diff}=\pi$ remains exact, so the drive matrix elements $\mel{j}{\hat H_d}{i}$ are nonzero only if $\ket{i}$ and $\ket{j}$ correspond to different parity.

On the other hand, some matrix elements such as $\mel{1}{\hat n}{0}$ decrease exponentially with increasing $E_\text{J}/E_L$ but retain a usable magnitude in the shallow double-well regime, since the barrier between the potential wells is not yet tall enough to fully suppress the wavefunction overlap [Fig.~\hyperref[fig:circuit_schematic]{\ref*{fig:circuit_schematic}(c)}]. The intermediate regime therefore offers the simultaneous availability of the single-excitation transitions inherited from the unimon regime and the multi-excitation transitions characteristic of the deeper double-well regimes.

This regime benefits the multiunimon in two ways. Firstly, it expands the set of capacitively connected states, providing more flexibility in choosing the eight computational eigenstates and assigning their logical labels to minimize leakage and other gate errors. Secondly, it enables additional native gates that are not accessible to devices using normal-mode encoding.

\subsection{Leakage-aware encoding}

Normal-mode encoding relies on a normal-mode description of the circuit, in which the potential is expanded around a single well-defined minimum and the resulting modes label the eigenstates. In the double-well regime this description breaks down. The potential has two minima, and the low-energy eigenstates are delocalized across both wells, so they no longer carry clean normal-mode labels. No natural mapping between the computational states $\ket{i j k}, \ i,j,k\in\{0,1\}$ and the eigenstates of Eq.~\eqref{eq:full_H} exists in this regime, but this is not a fundamental obstacle. The cube schematic of Fig.~\hyperref[fig:cube_schematic]{\ref*{fig:cube_schematic}(a)} is agnostic regarding the encoding, and any set of eight states with the required drive connectivity and selectivity can serve as the computational subspace. In this work, we restrict ourselves to eigenstates of the multiunimon Hamiltonian.

Ideally, the cube supports CC$\mathcal{R}$ rotations between computational states without activating any other transitions, such as off-resonant driving of spectator transitions in the computational subspace or transitions into the leakage subspace [Fig.~\hyperref[fig:cube_schematic]{\ref*{fig:cube_schematic}(a)}]. Although ac Stark errors can be mitigated with careful calibration~\cite{maurya_universal_2026}, designs that minimize these errors reduce the calibration burden and improve robustness. Incoherent error sources, such as dissipation from dielectric loss---estimated to be among the leading error sources in unimon circuits~\cite{hyyppa_unimon_2022}---also limit the gate fidelity of the multiunimon.

To identify the best encoding of the computational states onto the multiunimon eigenstates, we introduce a leakage-aware metric and a search procedure that uses it to find the encoding most likely to yield high-fidelity CC$\mathcal{R}$ gates. The metric accounts for leakage and crosstalk from off-resonant driving, ac Stark shifts, and decoherence contributions, while the search procedure efficiently explores the space of possible eight-state assignments.

The ideal metric would be the average gate infidelity (AGI) of the full native gate set. This is, however, not practical. The exact computation of the AGI is too expensive when evaluating many encoding candidates. We instead use a related metric based on pairwise infidelities calculated for every pair of states that involves at least one computational state. Of these pairs, only one corresponds to the desired CC$\mathcal{R}$ transition, while the remaining pairs are spectators that should ideally remain unaffected by the drive. Each pairwise infidelity accounts for off-resonant driving, ac Stark-induced phases, and decoherence on that pair. For each gate, the pairwise infidelities are combined into a single number by taking their root mean square. The twelve gate values are then averaged to obtain a single metric for the encoding candidate. In this work, the pairwise infidelities are calculated for gates with a rotation angle of $\pi$ and a gate length of \SI{30}{\nano\second}.

We emphasize that this pairwise metric is not a reliable estimator of the actual AGI. It serves only to rank encoding candidates within the optimization framework, and the AGI is computed separately for the optimized design. A detailed description of the metric is given in Appendix~\ref{app:fid_estimation}.

\subsection{Extended native gate set}

The native gate set of the multiunimon consists of twelve CC$\mathcal{R}$ gates and the corresponding virtual CCZ gates. The combination of increased drive connectivity in the shallow double-well regime and leakage-aware encoding, however, enables additional gates beyond this set. One particularly useful application is to choose an encoding that enables a single-pulse transition between the computational states $\ket{000}$ and $\ket{111}$. With a rotation angle of  $\pi/2$, this pulse implements a direct preparation of the maximally entangled GHZ state, defined as $\left( \ket{000} + \ket{111}\right)/\sqrt{2}$. More generally, the cube has four space diagonals connecting opposite vertices, and similar single-pulse transitions can be enabled along these diagonals to reduce the maximum graph distance between any two computational states from three to two. 
%\textcolor{orange}{(Fig.~X)}
In Section~\ref{sec:results}, we demonstrate a design that enables three of the four space diagonal transitions with high fidelity.

\begin{figure*}[t]
\includegraphics[scale=0.98]{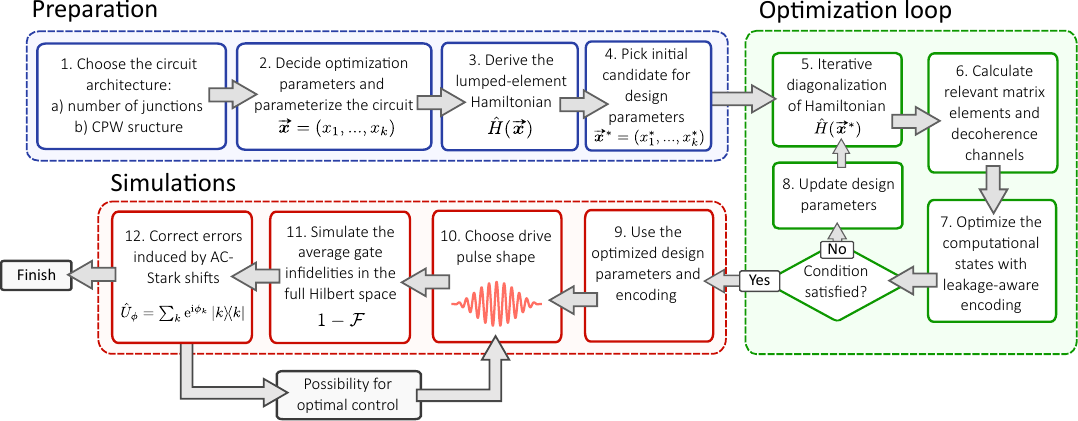}
\caption{\label{fig:opt_workflow}
Full workflow for multiunimon optimization and gate simulation. The workflow consists of three distinct parts: preparation, optimization, and simulation. In the preparation part, we choose the circuit architecture and model parameters, and derive the corresponding Hamiltonian. The optimization loop searches for the optimal parameters and encoding using the leakage-aware metric. Once a suitable candidate is found, the properties of the optimized multiunimon are passed to full simulations, where the phase-corrected average gate infidelities are calculated for a chosen pulse shape.
}
\end{figure*}

\section{\label{sec:optimization}Optimization workflow}

Having established the main design components of the multiunimon device, namely its circuit geometry and the encoding of the computational subspace, we lay out the general workflow used to optimize the device and compute the average gate infidelities. The workflow consists of three parts: preparation, optimization, and gate simulation.

\subsection{Preparation}

The workflow illustrated in Fig.~\ref{fig:opt_workflow} begins with the choice of device architecture. In this work, we focus on a single-junction architecture with branched CPW geometry, but the methods generalize to a broader set of architectures. For instance, the Hamiltonian of Eq.~\eqref{eq:full_H} describes any single-junction multiunimon device, and adding more junctions extends the framework to include additional nonlinear modes. The Hamiltonian is derived from the distributed-element Lagrangian by recasting it in a lumped-element form by using an auxiliary-mode decomposition of the CPW kernel, where the auxiliary modes correspond to bare CPW normal modes. For the branched CPW geometry used here, the kernel admits analytical expressions, providing a parameterization of the Hamiltonian in terms of the design parameters; however, this is not necessarily the case for more complicated CPW structures. A detailed derivation is given in Appendix~\ref{app:Derivation}.

\subsection{Optimization}

Optimizing the multiunimon requires two distinct sets of parameters, which we handle through a nested two-layer optimization loop. The outer layer optimizes the parameters describing the circuit design [Fig.~\hyperref[fig:circuit_schematic]{\ref*{fig:circuit_schematic}(a)}], while the inner layer optimizes the encoding of the computational subspace. Each iteration begins with a candidate set of design parameters, from which we construct the Hamiltonian.

For each candidate, we compute the low-energy eigenstates, energies, and relevant matrix elements using an iterative diagonalization scheme. The scheme starts with diagonalizing the nonlinear mode, which is then combined with a single auxiliary mode, jointly diagonalized, and truncated to retain only the low-energy subspace. The retained eigenstates are combined with the next auxiliary mode, and this is repeated until the desired mode cutoff is reached. Details are given in Appendix~\ref{app:NRG}. The effects of decoherence arising from dielectric loss and external flux noise, are then modeled from the eigenenergies and matrix elements, as described in Appendix~\ref{app:Decoherence}.

The inner layer of the optimization loop identifies the optimal leakage-aware encoding for a given set of design parameters. We use the leakage-aware metric described in Sec.~\ref{sec:encoding} and a search algorithm that efficiently prunes the search space and identifies promising encoding candidates (Appendix~\ref{app:cube_search}). The encoding minimizing the leakage-aware metric is recorded as the best encoding available for the current design parameter set. The outer layer then advances to a new set of design parameters, and the procedure repeats. The outer layer ranks candidate designs using the same leakage-aware metric.

The optimization combines continuous parameters in the outer layer with discrete parameters in the inner layer. The discrete nature of the encoding optimization makes the parameter landscape non-differentiable, so gradient-based methods are not effective. Instead, we use the differential evolution algorithm, a member of the genetic algorithm family, which is well-suited to gradient-free optimization with multiple minima and large parameter sets~\cite{weissler_enumeration_2024, garcia-azorin_robust_2026, cardenas-lopez_resilient_2025, menke_automated_2021}. The differential evolution algorithm is supplemented by a Nelder-Mead simplex for local polishing of the best candidate.

\subsection{Simulation}

With the optimization complete, we simulate the gate infidelities for the optimized design, using the eigenstates, energies, drive matrix elements, and decoherence rates obtained from the diagonalization. The simulations are carried out in a truncated Hilbert space spanning the states relevant to each gate, retaining all eigenstates that may affect the dynamics of the computational subspace. We neglect eigenstates with energy exceeding the sum of the highest-energy computational state and the highest drive frequency used for the CC$\mathcal{R}$ gates, since these states are far-detuned and unlikely to induce significant leakage or ac Stark shifts in the computational subspace.

As with single-qubit gates in superconducting qubits, the pulse shape can significantly affect the leakage and crosstalk~\cite{hyyppa_reducing_2024, wesdorp_mitigating_2026}. This is particularly true for the multiunimon, where a single drive line addresses all three qubits and the multimode structure produces a larger number of leakage channels than in typical superconducting qubit designs. Advanced pulse-shaping techniques can therefore improve the gate fidelities. In this work, however, we focus on design optimization and use sine-squared pulse shapes throughout.

Decoherence is modeled with a Lindblad master equation, from which we compute the AGIs using the definition of Ref.~\cite{pedersen_fidelity_2007}. For large Hilbert spaces, computing the full gate Liouvillian is computationally expensive. We instead use an approach that computes the AGI directly without constructing the full Liouvillian~\cite{cabrera_average_2007, orell_efficient_2026}, reducing the problem to evolving the generalized Pauli operators of the computational subspace under the master equation. 
%\textcolor{orange}{(Appendix~\ref{app:AGI_simulation})}.

To simulate the ability of virtual CCPHASE gates to correct ac Stark shifts~\cite{maurya_universal_2026}, we apply an additional unitary $\hat U_{\phi}=\text{diag}(\ee^{\ii\phi_1}, ..., \ee^{\ii\phi_8})$ representing the above-discussed frame update. The phases $\phi_k$ are optimized to minimize the AGI, yielding the optimal phase corrections. 
%\textcolor{orange}{(Appendix~\ref{app:ac_stark})}.

\section{\label{sec:results}Results}

\subsection{The optimized design and encoding}

\begin{figure*}[t]
\includegraphics[scale=0.92]{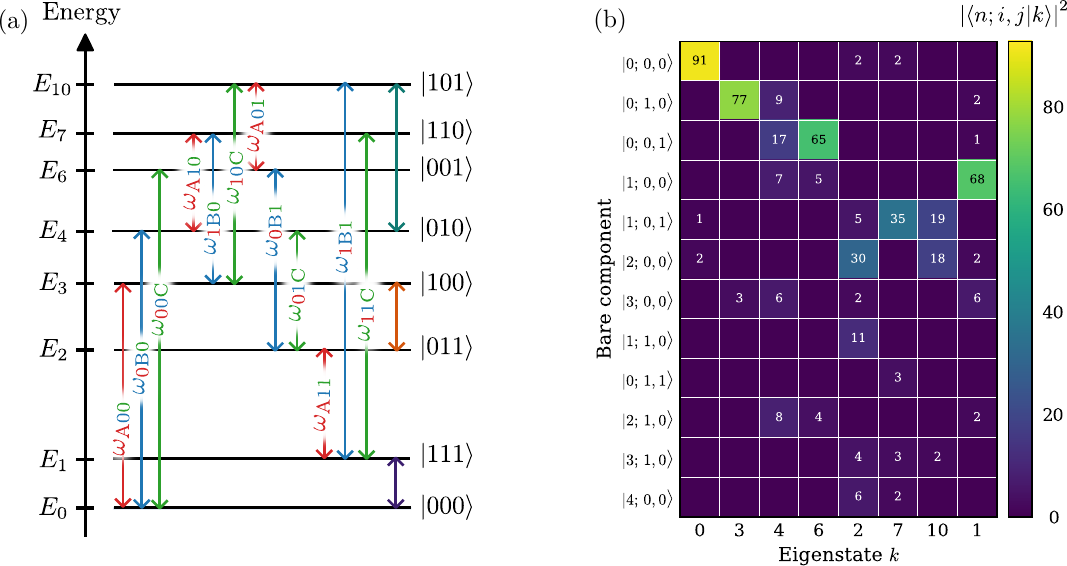}
\caption{\label{fig:energy_fig}
(a) Energy-level diagram of the computational states, drawn to scale, with the native gate transition frequencies indicated. The three rightmost transitions are the extended gate-set transitions enabled by the leakage-aware encoding. (b)~Bare-component contributions to each computational eigenstate. The bare components are the product states of the nonlinear and linear auxiliary modes [Eqs.~\eqref{eq:H_JJ} and~\eqref{eq:H_CPW}], denoted as $\ket{n;i,j}$, where $n$ is the excitation number of the nonlinear mode and \{$i$,~$j$\} are the excitation numbers of the two lowest-frequency auxiliary modes. The overlaps $\abs{\braket{n;i,j|k}}^2$ with each computational eigenstate $\ket{k}$ are shown as percentages, with values below 1\% omitted.
}
\end{figure*}

\begin{figure*}[t]
\includegraphics[scale=0.542]{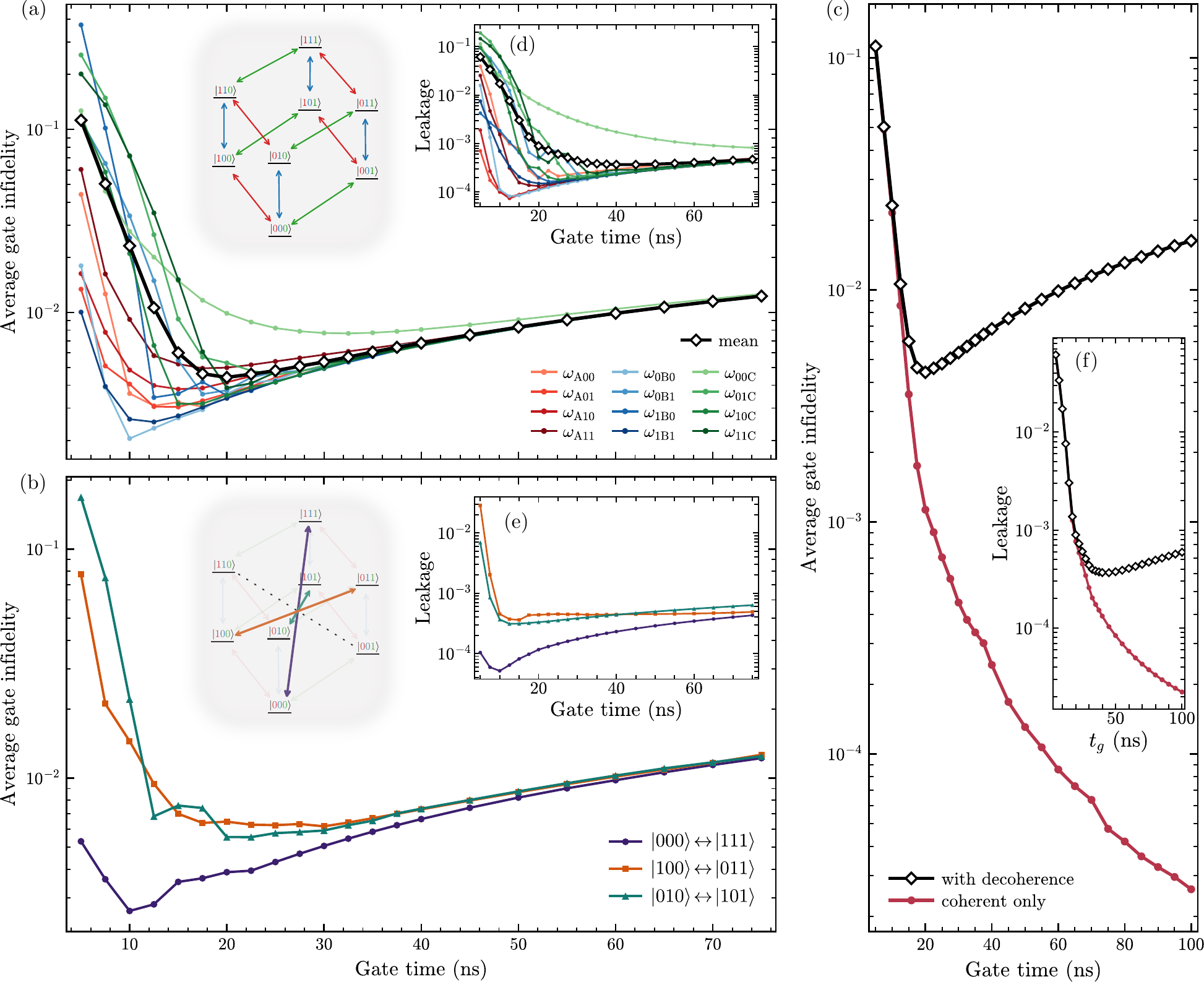}
\caption{\label{fig:result_AGI} 
(a)–(c) Simulated average gate infidelities (AGI) for $\pi$ rotations as functions of gate time. (a) AGIs for the indicated native CCNOT gates each corresponding to an edge of the cube shown in the inset; the mean across all twelve CCNOTs is shown as a black curve. (b) AGIs for the three transitions corresponding to the indicated diagonals of the cube shown in the inset. (c) Mean AGI over the twelve CCNOT gates considered in (a) as a function of gate time, with (black color) and without (red color) incoherent errors. The black curve corresponds to that in (a), and the red curve includes only coherent errors such as leakage (see inset) and crosstalk from off-resonant driving. (d)–(f) Population leakage outside the computational subspace, for the rotations considered in (a)–(c), respectively, as a function of gate time. All simulations use the optimized design and encoding of Tables~\ref{tab:design} and~\ref{tab:comp_energies}, and assume correction of deterministic phase errors from ac Stark shifts.
}
\end{figure*}

We illustrate the capabilities of the multiunimon by presenting optimized results for an example design. Table~\ref{tab:design} summarizes the design parameters identified by the outer layer of the optimization workflow of Section~\ref{sec:optimization}. The leakage-aware computational state mapping from the inner layer, along with the corresponding eigenstate energies and relaxation times, are shown in Table~\ref{tab:comp_energies}.

\begin{table}[b]
  \caption{Design and the main noise parameters of the multiunimon circuit. Positions are measured from the centre of the device along the corresponding branch as shown in Fig.~\hyperref[fig:circuit_schematic]{\ref*{fig:circuit_schematic}(a)}.}
  \label{tab:design}
  \begin{ruledtabular}
    \begin{tabular}{l c S[table-format=-3.3] l}
      Parameter & Symbol & {Value} & Unit \\ \hline
      Josephson energy\footnote{Free parameter in the design optimization.}
                                    & $E_\text{J}$ &  25.651 & GHz \\
      Junction capacitance\footnotemark[1]
                                    & $C_\text{J}$ &   2.418 & fF  \\ \hline
      $x$-branch half-length\footnotemark[1]
                                    & $l_x$        &   3.607 & mm  \\
      $y$-branch half-length\footnotemark[1]
                                    & $l_y$        &   4.783 & mm  \\
      Junction position\footnotemark[1]
                                    & $x_\text{J}$ &  -1.607 & mm  \\
      $x$-branch intersection point\footnotemark[1]
                                    & $x_\text{b}$ &   0.930 & mm  \\
      $y$-branch intersection point\footnotemark[1]
                                    & $y_\text{b}$ &   1.728 & mm  \\
      Drive position\footnotemark[1]
                                    & $x_\text{d}$ &   1.389 & mm  \\ \hline
      Inductance per length\footnote{Depends on the values of $Z_v$ and $v$.}         & $L_\ell$     &   1.861 & \textmu H/m \\
      Capacitance per length\footnotemark[2]        & $C_\ell$     &  38.302 & pF/m \\
      Drive capacitance             & $C_\text{d}$ &   0.083 & fF  \\ \hline
      Characteristic impedance\footnotemark[1]
                                    & $Z_c$        & 220.420 & $\Omega$ \\
      Phase velocity\footnotemark[1]                & $v$          &   0.119 & m/ns \\
      Josephson-to-inductive ratio\footnotemark[2]                & $E_\text{J}/E_L$          &   1.670 & --- \\ \hline
      Dielectric quality factor
                                    & $Q_\text{diel}$        & {$3.5 \times 10^5$} & --- \\
      Flux noise density at \SI{1}{\hertz}               & $A_{\Phi_{\text{diff}}}$          &   15.0 & $\upmu \Phi_0$ \\
      Bath temperature                & $T$          &   25.0 & mK \\
    \end{tabular}
  \end{ruledtabular}
\end{table}

The mapping between eigenstates and computational states departs from normal-mode encoding. Most strikingly, the first excited eigenstate is assigned to the computational state $\ket{111}$. In the unimon~\cite{hyyppa_unimon_2022}, the ground and first excited states define the two qubit levels; in the multiunimon, the same two eigenstates instead define $\ket{000}$ and $\ket{111}$, the all-ground and all-excited computational states. Figure~\hyperref[fig:energy_fig]{\ref*{fig:energy_fig}(a)} shows the energy-level structure of the computational subspace, with all native CC$\mathcal{R}$ transition frequencies drawn to scale. Compared to the trimon~\cite{roy_programmable_2020, maurya_universal_2026}, these transition frequencies span a substantially wider range, from 3 to \SI{13.5}{\giga\hertz}, with a minimum separation of roughly \SI{370}{\mega\hertz} (Table~\ref{tab:transitions}). This reduces crowding in the transition frequencies, a known limitation of the trimon, allowing shorter pulses to drive each transition without exciting its neighbors and thereby enabling faster gates.

In this case, the gates are protected against unwanted off-resonant transitions within the computational subspace by spectral separation alone. Leakage out of the computational subspace is less straightforward to assess, because a transition into the leakage subspace can lie close to resonance with the drive and still have only small leakage contribution if its drive coupling is weak. For example, the rotation between $\ket{000}$ and $\ket{100}$ at frequency $\omega_{\text{A}00}/2\pi$ is separated by only about \SI{1}{\mega\hertz} from a transition connecting $\ket{001}$ to the leakage subspace. The drive coupling to that leakage transition is two orders of magnitude weaker compared to the target transition, which keeps the resulting leakage at a manageable level.

This is one reason we use the leakage-aware metric. Unlike anharmonicities and cross-Kerr shifts, which characterize only the spectrum, it incorporates the drive coupling of each transition alongside with the necessary detunings to estimate the leakage.

More broadly, these features illustrate the central advantage of leakage-aware encoding: the computational subspace is chosen to optimize performance rather than to follow a predetermined labeling. The encoding flexibility of this design arises from its Josephson-to-inductive energy ratio, which places the potential of the nonlinear mode in the shallow double-well regime [Fig.~\hyperref[fig:circuit_schematic]{\ref*{fig:circuit_schematic}(c)}] and provides the richer drive connectivity discussed in Sec~\ref{sec:encoding}. As shown in Fig.~\hyperref[fig:energy_fig]{\ref*{fig:energy_fig}(b)}, many of the computational states are superpositions of multiple bare-mode excitations, while still respecting the parity conservation enforced by the symmetric potential at $\varphi_\text{diff} = \pi$.

\begin{table}
  \caption{Eigenstate-to-computational-state mapping, energies relative to the ground state, and relaxation times $T_1$ of the computational states which are mainly limited by dielectric losses (see Table~\ref{tab:design} and Appendix~\ref{app:Decoherence})}
  \label{tab:comp_energies}
  \begin{ruledtabular}
    \begin{tabular}{c c d{2.3} d{2.1}}
      \makecell{Eigenstate \\ $k$} &
      \makecell{Computational \\ state} &
      \multicolumn{1}{c}{\makecell{$E_k - E_0$ \\ (GHz)}} &
      \multicolumn{1}{c}{\makecell{$T_1$ \\ (\textmu s)}} \\ \hline
       0 & $\lvert 000 \rangle$ &  0.000 & 711.7 \\
       1 & $\lvert 111 \rangle$ &  1.803 & 16.4 \\
       2 & $\lvert 011 \rangle$ &  5.724 &  6.6 \\
       3 & $\lvert 100 \rangle$ &  8.114 &  6.1 \\
       4 & $\lvert 010 \rangle$ & 10.014 &  4.4 \\
       6 & $\lvert 001 \rangle$ & 12.226 &  4.3 \\
       7 & $\lvert 110 \rangle$ & 13.537 &  3.4 \\
      10 & $\lvert 101 \rangle$ & 15.301 &  3.1 \\
    \end{tabular}
  \end{ruledtabular}
\end{table}

\subsection{Native gate fidelities}

\begin{table}[b]
  \caption{Transition frequencies $f$ and optimal average gate (in)fidelities, together with the corresponding gate times, for the 12 native CC$\mathcal{R}$ rotations and three cube space-diagonal rotations. All values correspond to $\pi$ rotations.}
  \label{tab:transitions}
  \begin{ruledtabular}
    \begin{tabular}{ccS[table-format=2.3]ccc}
      Transition & Symbol & {$f$ (GHz)} & $1-\bar{\mathcal{F}}$ & $\bar{\mathcal{F}}$ (\%) & $t_g$ (ns) \\
      \hline
      $\ket{000} \leftrightarrow \ket{100}$ & $\omega_{\text{A}00}$ &  8.114 & $3.0 \times 10^{-3}$ & 99.70 & 17.5 \\
      $\ket{010} \leftrightarrow \ket{110}$ & $\omega_{\text{A}10}$ &  3.523 & $3.8 \times 10^{-3}$ & 99.62 & 15.0 \\
      $\ket{001} \leftrightarrow \ket{101}$ & $\omega_{\text{A}01}$ &  3.075 & $3.0 \times 10^{-3}$ & 99.70 & 15.0 \\
      $\ket{011} \leftrightarrow \ket{111}$ & $\omega_{\text{A}11}$ &  3.921 & $5.0 \times 10^{-3}$ & 99.50 & 17.5 \\
      \hline
      $\ket{000} \leftrightarrow \ket{010}$ & $\omega_{0\text{B}0}$ & 10.014 & $2.0 \times 10^{-3}$ & 99.80 & 10.0 \\
      $\ket{100} \leftrightarrow \ket{110}$ & $\omega_{1\text{B}0}$ &  5.423 & $3.4 \times 10^{-3}$ & 99.66 & 12.5 \\
      $\ket{001} \leftrightarrow \ket{011}$ & $\omega_{0\text{B}1}$ &  6.502 & $3.6 \times 10^{-3}$ & 99.64 & 17.5 \\
      $\ket{101} \leftrightarrow \ket{111}$ & $\omega_{1\text{B}1}$ & 13.498 & $2.5 \times 10^{-3}$ & 99.75 & 12.5 \\
      \hline
      $\ket{000} \leftrightarrow \ket{001}$ & $\omega_{00\text{C}}$ & 12.226 & $7.7 \times 10^{-3}$ & 99.23 & 32.5 \\
      $\ket{100} \leftrightarrow \ket{101}$ & $\omega_{10\text{C}}$ &  7.188 & $3.2 \times 10^{-3}$ & 99.68 & 15.0 \\
      $\ket{010} \leftrightarrow \ket{011}$ & $\omega_{01\text{C}}$ &  4.290 & $4.6 \times 10^{-3}$ & 99.54 & 25.0 \\
      $\ket{110} \leftrightarrow \ket{111}$ & $\omega_{11\text{C}}$ & 11.734 & $3.9 \times 10^{-3}$ & 99.61 & 20.0 \\
      \hline
      $\ket{000} \leftrightarrow \ket{111}$ & --- &  1.803 & $2.6 \times 10^{-3}$ & 99.74 & 10.0 \\
      $\ket{100} \leftrightarrow \ket{011}$ & --- &  2.389 & $6.2 \times 10^{-3}$ & 99.38 & 30.0 \\
      $\ket{010} \leftrightarrow \ket{101}$ & --- &  5.287 & $5.5 \times 10^{-3}$ & 99.45 & 22.5 \\
    \end{tabular}
  \end{ruledtabular}
\end{table}

We assess the performance of the multiunimon by simulating the average gate infidelity $1 - \bar{\mathcal{F}}$, with fidelity $\bar{\mathcal{F}}$ defined as in Ref.~\cite{pedersen_fidelity_2007}. Figure~\hyperref[fig:result_AGI]{\ref*{fig:result_AGI}(a)} shows the infidelity of each native CC$\mathcal{R}$ rotation as a function of gate time, at a rotation angle of $\pi$. Combined with virtual Z gates, these CC$\mathcal{R}(\pi)$ rotations implement CCNOT gates. At a uniform gate time of \SI{20}{\nano\second}, the mean infidelity across the twelve gates is $4.4 \times 10^{-3}$ ($\bar{\mathcal{F}}= 99.56\%$), ranging from $2.0 \times 10^{-3}$ ($99.80\%$) for the best gate ($\ket{000} \leftrightarrow \ket{010}$, at a \SI{10}{\nano\second} gate time) to $7.7 \times 10^{-3}$ ($99.23\%$) for the worst. The mean improves slightly when the gate time is individually optimized for each gate (Table~\ref{tab:transitions}).

The optimal gate times are fast, on par with single-qubit gates that use advanced pulse-shaping techniques~\cite{hyyppa_reducing_2024}. These short gate times reflect the dominance of decoherence over coherent control errors. Since longer gates accumulate more decoherence, the optimal gate is pushed toward short pulses. The dominant decoherence channel is dielectric loss (Appendix~\ref{app:Decoherence}), which gives $T_1$ relaxation times ranging from \SI{711.7}{\micro\second} for $\ket{000}$ to \SI{3.1}{\micro\second} for $\ket{101}$. Table~\ref{tab:comp_energies} shows an inverse relation between state energy and $T_1$, which we attribute to the stronger coupling and the larger number of available decay channels at higher energies. The stronger coupling follows from the charge zero-point fluctuations, which scale roughly as $\propto \sqrt{\omega}$ in the harmonic-oscillator basis, giving a dielectric relaxation rate that grows linearly with transition frequency.

To isolate the coherent-error contribution, we simulate the mean AGI for the native CCNOT gates with decoherence excluded [Fig.~\hyperref[fig:result_AGI]{\ref*{fig:result_AGI}(c)}]. The performance improves substantially, with the mean infidelity dropping below $10^{-3}$ (99.9\%) at a \SI{20}{\nano\second} gate time and below $10^{-4}$ (99.99\%) around \SI{60}{\nano\second}. This demonstrates the performance attainable if the dominant noise sources, particularly dielectric loss, are reduced through improved materials, fabrication processes, and design~\cite{bal_systematic_2024, megrant_planar_2012, place_new_2021, bland_2d_2025, lahtinen_effects_2020}. Notably, these results assume a relatively low dielectric quality factor, $Q_\text{diel}=3.5 \times 10^5$~\cite{hyyppa_unimon_2022}, so improvements in dielectric quality would directly raise the achievable fidelities. The relatively high characteristic impedance $Z_c$, on the other hand, suppresses $1/f$ flux noise (Appendix~\ref{app:Decoherence}). Although the characteristic impedance $Z_c \approx \SI{220}{\ohm}$ is higher than what is typical for CPWs, experimental realizations at these and higher impedances have been demonstrated. Approaches for achieving high impedance, including the superinductor regime~\cite{bell_quantum_2012, masluk_microwave_2012}, include devices using large geometric inductances~\cite{peruzzo_surpassing_2020, medahinne_magnetic-field-tolerant_2025}, high kinetic inductances~\cite{niepce_high_2019, frasca_nbn_2023, hazard_nanowire_2019, grunhaupt_granular_2019}, and Josephson junction arrays~\cite{ranni_high_2023}.

Importantly, we used sine-squared drive pulses in the simulations, leaving substantial room for advanced pulse-shaping techniques to reduce the already low coherent error rate further. For instance, sophisticated DRAG techniques such as those introduced in Ref.~\cite{hyyppa_reducing_2024} may help to suppress leakage to allow faster gates, which in turn reduces the exposure time to decoherence and raises the gate fidelity. Such pulses may be particularly useful for the multiunimon, since they can be designed to suppress several leakage channels simultaneously.

\subsection{Extended gate set fidelities}

The leakage-aware encoding allows computational subspaces that depart from normal-mode encoding, yielding the high-fidelity native CCNOT gates described above. This gate set can be extended to include microwave-driven transitions beyond the cube edges shown in Fig.~\ref{fig:cube_schematic}. These additional transitions connect states differing in all three logical bits, such as the $\ket{000} \leftrightarrow \ket{111}$ transition, corresponding to the space diagonals of the cube. There are four such space diagonals. For this design, three are accessible with $\pi$-rotation infidelities below $7 \times 10^{-3}$ ($99.3\%$), reaching $2.6 \times 10^{-3}$ ($99.74\%$) for the $\ket{000} \leftrightarrow \ket{111}$ transition at a \SI{10}{\nano\second} gate time [Fig.~\hyperref[fig:result_AGI]{\ref*{fig:result_AGI}(b)}]. The fourth diagonal, $\ket{001} \leftrightarrow \ket{110}$, does not reach high fidelity in this design, owing to its weak coupling to the drive. There is no fundamental obstacle, however, to accessing all four diagonals in a single design. The frequencies and infidelities of these transitions are detailed in Table~\ref{tab:transitions}.

%\textcolor{orange}{Such transitions are not available as native single-step gates in normal-mode encoding.}
The $\ket{000} \leftrightarrow \ket{111}$ transition enables single-step preparation of the GHZ state, and more generally these diagonals reduce the maximum graph distance between the computational states from three to two, lowering the circuit depth required to connect arbitrary multi-qubit states. The complete native gate set thus comprises the twelve CC$\mathcal{R}$ gates, the virtual CCPHASE family, and three cube-diagonal transitions.

\section{\label{sec:summary_discussions}Summary and discussions}

We have introduced the multiunimon, a three-qubit device based on a unimon-type circuit that consists of only a single Josephson junction embedded in a CPW structure. To improve the multi-qubit performance, we extended the original unimon design by adding a branch to the CPW structure and operating in the shallow double-well regime. These changes, combined with a leakage-aware encoding of the computational subspace, support twelve generalized CCNOT gates with an average simulated infidelity of $4.4 \times 10^{-3}$ at a \SI{20}{\nano\second} gate time, limited primarily by dielectric losses. Our simulations including only coherent errors suggest CCNOT infidelities below $10^{-4}$ at gate times around 60 ns. 

The largest performance improvements are likely to come from addressing the incoherent error sources, particularly dielectric loss. Raising the dielectric quality factor $Q_\text{diel}$ above $10^6$, which is within experimental reach~\cite{megrant_planar_2012, place_new_2021}, would already yield substantial gains. Beyond these, the coherent error rate can be lowered with advanced pulse-shaping techniques such as DRAG~\cite{hyyppa_reducing_2024, motzoi_simple_2009} or methods designed to reduce crosstalk~\cite{wesdorp_mitigating_2026}. These are especially relevant for a device like the multiunimon, which relies on a single drive line to control multiple qubits. In particular, calibrated off-resonant $\pi/2$ pulses~\cite{wesdorp_mitigating_2026} combined with virtual Z gates may provide additional flexibility, especially if a single multiunimon encodes more than three qubits.

In addition to achieving high-fidelity native CC$\mathcal{R}$ gates, the leakage-aware encoding enables extended transitions that change the state of each qubit. The connectivity may be increased further with Raman-based gates, recently demonstrated in Ref.~\cite{maurya_universal_2026}, which use a mediator state in the opposite parity sector to connect computational states within the same parity sector. On the cube (Fig.~\ref{fig:cube_schematic}), these extended transitions correspond to the face diagonals, enabling gates such as the controlled-$\sqrt{i\text{SWAP}}$ and controlled-$\sqrt{i\text{bSWAP}}$, which combined with virtual Z gates can realize the Fredkin gate. Together, the twelve CC$\mathcal{R}$ gates (cube edges), the four space diagonals, and the Raman-based face diagonals provides a computational subspace with all-to-all connectivity between the eight states.

All-to-all connectivity is especially valuable if the eight-state subspace is used as a single qudit. Superconducting qudits are typically limited by selection rules that restrict connectivity to a ladder-like structure~\cite{li_universal_2025, wang_high-_2025}, requiring carefully optimized pulse sequences for arbitrary qudit control. The multiunimon may instead provide all-to-all connectivity using a single pulse to connect most qudit levels, while the rest may be connected with Raman-based pulses.% for transitions between opposite parity sectors and individual two-tone Raman pulses for transitions within the same sector.

Dispersive readout has been demonstrated for the unimon~\cite{hyyppa_unimon_2022} and is the standard approach for trimon devices~\cite{roy_implementation_2017, roy_multimode_2018, roy_programmable_2020, maurya_universal_2026}. In the trimon, however, the symmetric mode structure produces only weakly distinguishable dispersive shifts across the computational states, leading to insufficient single-shot separability that is often addressed by performing multiple readout rounds~\cite{roy_programmable_2020}. A similar dispersive approach can be implemented for the multiunimon, which is not constrained by such mode symmetries. The leakage-aware encoding further implies that the computational states are likely to have substantially different frequencies, improving their separability. In addition, the readout resonator can be coupled at any single point along the CPW structure, or at multiple points, providing flexibility to engineer the dispersive shifts for efficient readout. On the other hand, the richer connectivity through the charge matrix elements can constrain the choice of the readout resonator frequency since several transitions are susceptible to Purcell decay. We also note that the trimon inherits the cosine-shaped potential of the transmon and is therefore susceptible to measurement-induced transitions to free-particle-like states, which are known to degrade readout fidelity and its quantum non-demolition (QND) character~\cite{lescanne_escape_2019, shillito_dynamics_2022, cohen_reminiscence_2023}. Like the unimon, the multiunimon has a fully bounded potential and is therefore protected against such transitions~\cite{hyyppa_unimon_2022}.

Scaling the multiunimon to large qubit numbers can proceed along two fronts: (i) encoding more qubits within a single device and (ii) enabling entangling gates between devices. For (i), a single multiunimon encodes more qubits, with the native rotations generalizing to $\text{C}^{n-1}\mathcal{R}$ for $n$ qubits. With careful optimization of the design, encoding, and pulse shaping, four- or even five-qubit multiunimon devices may be feasible, but beyond this, frequency crowding is likely to become a serious bottleneck. Alternative circuit architectures with additional junctions or different CPW geometries possibly provide an easier pathway to larger qubit numbers. For (ii), multiple multiunimon devices are coupled, for example through capacitive coupling, with inter-device entangling gates realized by the cross-resonance technique~\cite{chow_simple_2011, rigetti_fully_2010, paraoanu_microwave-induced_2006}. This resembles the proposals for coupling trimon blocks into larger processors~\cite{roy_multimode_2018, roy_programmable_2020, hazra_engineering_2020} and for coupling unimon qubits~\cite{hyyppa_unimon_2022}.

Altogether, these results establish the multiunimon as a single-junction platform for microwave-driven native multi-qubit gates with all-to-all connectivity. The leakage-aware encoding, which selects the computational subspace directly from the physical eigenstates, is central to enabling high-fidelity gate operations in parameter regimes where conventional normal-mode encoding would fail. With further improvements in materials, circuit design, and pulse shaping, we envision the multiunimon as a platform for small-scale quantum processing units, with the potential to scale further once suitable inter-device entangling mechanisms are developed.

%However, due to the large search space and the imperfect cost function, better-performing designs likely exist within this architecture but were not identified by our search.

%\textcolor{orange}{Maybe we should say something about the fact that the encoding in question can lead to highly correlated errors such as correlated relaxation $\ket{111} \rightarrow \ket{000}$ that would be problematic for typical error-correction codes that assume local errors. However, our design choise was to optimize for gate fidelity and computational state connectivity, and criteria could be changed to look for more error-correction friendly encodings.}

\section*{Acknowledgments}
We thank Vasilii Vadimov, Heikki Suominen, and Wallace Teixeira for useful discussions. We acknowledge financial support from the Research Council of Finland through the Academy Professors's project Autonomous Quantum Machines (Grant No. $369679$), from the European Research Council under the Advanced Grant ConceptQ (No. $101053801$), from Jane and Aatos Erkko Foundation through the SystemQ project, from Novo Nordisk Foundation through the project Sustainable Quantum Computers (No. NNF$25$OC$0103322$), and from the H2Future project through the Research Council of Finland (Grant. No. $352788$) and the University of Oulu. We also thank the Vilho, Yrjö and Kalle Väisälä Foundation of the Finnish Academy of Science and Letters for funding.

\appendix

\section{Derivation of auxiliary-mode Hamiltonian} \label{app:Derivation}

The multiunimon circuit consists of two intersecting CPW lines with uniform capacitance and inductance per unit length, $C_\ell$ and $L_\ell$. The line along the $x$ axis carries the flux field $\phi_x(x,t)$ over $x \in \left[-l_x, l_x \right]$, and the line along the $y$ axis carries $\phi_y(y,t)$ over $y \in \left[-l_y, l_y \right]$. Each line is grounded at both ends,
\begin{align}
    \phi_x(\pm l_x, t) = \phi_y(\pm l_y, t) = 0.
    \label{eq:grounding}
\end{align}
The flux field $\phi_x(x,t)$ is interrupted by a Josephson junction at location $x_\text{J}$, which allows the flux field to be discontinuous across the junction. We denote this jump by 
\begin{align}
    \Psi(t) \equiv \phi_x(x_\text{J}^+, t) - \phi_x(x_\text{J}^-, t).
    \label{eq:Psi_JJ}
\end{align}
Two other points on the circuit play a special role: (i) the branching point $(x_\text{b}, y_\text{b})$, where the two lines meet and (ii) the point $x_\text{d}$, where the drive line couples to the CPW. We place this coupling on the line along the $x$ axis, although any point on the line along the $y$ axis may work equally well.

The classical Lagrangian of the circuit can be expressed as
\begin{align}
    L &= L_\text{CPW} + L_\text{JJ} + L_\text{d},
\end{align}
where each part is defined as 
\begin{align}
    L_\text{CPW} &= \sum_{\alpha \in \{ x,y \}}\int_{-l_\alpha}^{l_\alpha} \mathrm{d}u \biggl\{ \frac{C_\ell}{2} \left[ \partial_t \phi_\alpha(u,t) \right]^2 \notag \\
    &\hspace{7.2em} -\frac{1}{2L_\ell}\bigl[\partial_u\phi_\alpha(u,t)\bigr]^2 \biggr\}, \label{eq:L_CPW} \\      
    L_\text{JJ}  &= \frac{C_\text{J}}{2} \dot{\Psi}^2(t) 
                   + E_\text{J} \cos \left[ \frac{2\pi}{\Phi_0} 
                   \Bigl( \Psi(t) - \Phi_\text{diff} \Bigr) \right], \label{eq:L_JJ} \\
    L_\text{d}   &= \frac{C_\text{d}}{2} \Bigl[ \dot{\phi}_x(x_\text{d},t) - V_\text{d}(t) \Bigr]^2. \label{eq:L_D}
\end{align}
The integral in Eq.~\eqref{eq:L_CPW} runs over the continuous parts of both lines, and leaves out the discontinuous junction point at $x_\text{J}$. The junction instead enters through the separate term $L_\text{JJ}$.

Note that many of the steps that follow mirror the more detailed treatment in the appendix of Ref.~\cite{tuohino_multimode_2024}, which therefore serves as a useful supplement.

\subsection{Equations of motion}

The equations of motion for the multiunimon circuit follow from the Euler--Lagrange equations. Away from the special locations, the flux field $\phi_\alpha$ obeys the wave equation
\begin{align}
    C_\ell \partial_t^2 \phi_\alpha(u,t) - \frac{1}{L_\ell} \partial_u^2\phi_\alpha(u,t) = 0, \quad \alpha \in \{ x,y \}.
    \label{eq:wave_equation}
\end{align}
The distinctive behavior of the circuit is encoded in the matching conditions at its special locations. In addition to the grounding at the ends of each line, the circuit has three such locations. The first is the Josephson junction, where the flux discontinuity $\Psi$ satisfies
\begin{align}
    C_\text{J} \ddot \Psi + I_c \sin \! \left[ \frac{2 \pi}{\Phi_0} \Bigl( \Psi - \Phi_\text{diff} \Bigr) \right] = \frac{1}{L_\ell} \partial_x\phi_x\big|_{x=x_\text{J}}.
    \label{eq:JJ_EoM}
\end{align}
Equation~\eqref{eq:JJ_EoM} couples the junction dynamics to the flux field of the CPW. The spatial derivative on its right-hand side is single-valued because the current is continuous across the junction,
\begin{align}
\jump{\partial_x \phi_x}{x_\text{J}} = 0,
\label{eq:current_JJ}
\end{align}
where $\jump{f}{c} = f(c^+) - f(c^-)$ denotes a jump in $f$ across the point $c$. The second special location is the branching point. There the flux takes a single value, so that
\begin{align}
\phi_x(x_\text{b}, t) = \phi_y(y_\text{b}, t),
\label{eq:branch_flux_condition}
\end{align}
and the currents from the branches sum to zero,
\begin{align}
\jump{\partial_x \phi_x}{x_\text{b}} + \jump{\partial_y \phi_y}{y_\text{b}} = 0,
\label{eq:branch_current_condition}
\end{align}
The third special location is the drive. However, because its coupling to CPW is weak, we neglect it at this stage of the derivation.

\subsection{Auxiliary-mode representation for the flux field}

Here, we focus our attention to the CPW and junction parts of the Lagrangian and express the CPW flux fields $\phi_x(x,t)$ and $\phi_y(y,t)$ in terms of $\Psi(t)$. Working in the frequency domain, where $\tilde{f}(\omega) = \int_{-\infty}^{\infty} f(t)\, 
\ee^{\ii\omega t}\, \mathrm{d}t$, we obtain the solutions
\begin{align}
    \tilde{\phi}_{\alpha}(u,\omega) &= -F_{\alpha}(u,\omega) 
    \tilde{\Psi}(\omega), \ \ \alpha \in \{x, y\},
\end{align}
where $F_{\alpha}(u,\omega) = N_{\alpha}(u,\omega) / D(\omega)$ and the explicit piecewise definitions of the kernels $N_\alpha$ and $D$ are given in Appendix~\ref{app:kernel_defs}.

We apply a variant of the Mittag-Leffler pole expansion,
\begin{align}
    F_{\alpha}(u, \omega) &\approx F_{\alpha}(u, 0) 
    + \frac{1}{2} \partial_\omega^2 F_{\alpha}(u, \omega)
    \big|_{\omega=0} \omega^2 \notag \\
    &+ \sum_{m=1}^M \frac{2f_{\alpha,m}(u)}{\Omega_m} \biggl[ 
    \frac{\Omega_m^2}{\omega^2 - \Omega_m^2} + 1 
    + \frac{\omega^2}{\Omega_m^2} \biggr],
\end{align}
where $\Omega_m$ and $f_{\alpha,m}$ are the location and residue of pole $m$, respectively \cite{tuohino_multimode_2024}. The residues are computed efficiently via
\begin{align}
    f_{\alpha,m}(u) = \frac{N_{\alpha}(u,\omega)}
    {\partial_\omega D(\omega)}\bigg|_{\omega=\Omega_m}.
\end{align}
By combining all $m$-dependent terms, this reduces to
\begin{align}
    F_{\alpha}(u, \omega) &\approx F_{\alpha}(u, 0) 
    + \frac{1}{2} \partial_\omega^2 F_{\alpha}(u, \omega)
    \big|_{\omega=0} \notag \\
    &+ \sum_{m=1}^M \frac{2f_{\alpha,m}(u)}{\Omega_m^3} 
    \frac{\omega^4}{\omega^2 - \Omega_m^2}.
\end{align}
The time-nonlocal term is eliminated by introducing 
a set of auxiliary modes
\begin{align} \label{eq:chi_kernel}
    \tilde{\chi}_m(\omega) &= \xi_m \frac{\omega^2}{\omega^2 - \Omega_m^2} 
    \tilde{\Psi}(\omega),
\end{align}
with mode-dependent coefficients $\xi_m$, which are defined in Eq.~\eqref{eq:xi_coeff}. The flux field may then be expressed in terms of 
the auxiliary modes, taking the form
\begin{align} \label{eq:aux_mode_flux_field}
    \tilde{\phi}_{\alpha}(u,\omega) &= -\biggl[ F_{\alpha}(u, 0) + \frac{1}{2} \partial_\omega^2 F_{\alpha}(u, \omega) \big|_{\omega=0} \omega^2 \biggr] \tilde{\Psi}(\omega) \notag \\
    &-\sum_{m=1}^M \frac{2f_{\alpha,m}(u)}{\Omega_m^3} 
    \frac{\omega^2}{\xi_m} \tilde{\chi}_m(\omega).
\end{align}

\subsection{Auxiliary-mode Lagrangian}

Using Eq.~\eqref{eq:aux_mode_flux_field}, the right-hand side of Eq.~\eqref{eq:JJ_EoM} becomes

\begin{align} \label{eq:JJ_EoM_RHS}
    \frac{1}{L_\ell}\partial_x \phi_x(x,t)\big|_{x=x_\text{J}} \!\!
    = -K\Psi(t) + \frac{K''}{2} \ddot{\Psi}(t) 
    + \sum_{m=1}^M \xi_m \ddot{\chi}_m(t),
\end{align}
where we have defined
\begin{align}
    K    &\equiv \frac{1}{L_\ell}
    \partial_x F_{x}(x,0)\big|_{x=x_\text{J}}, \\
    K''  &\equiv \frac{1}{L_\ell}
    \partial_x \partial_\omega^2 F_{x}(x,\omega)
    \big|_{\substack{\omega=0 \\ x=x_\text{J}}}, \\
    \xi_m &\equiv \sqrt{\frac{1}{L_\ell}
    \frac{2\partial_x f_{x,m}(x)\big|_{x=x_\text{J}}}{\Omega_m^3}}. \label{eq:xi_coeff}
\end{align}
Both $\partial_x F_x$ and $\partial_x f_{x,m}$ are continuous at $x_\text{J}$ by Eq.~\eqref{eq:current_JJ}. The equation of motion in Eq.~\eqref{eq:JJ_EoM} can then be written as
\begin{align} \label{eq:JJ_EoM_aux}
    C_\Psi \ddot{\Psi}(t) + \frac{\Psi(t)}{L_\Psi} 
    + I_\text{c} \sin \biggl[ \frac{2\pi}{\Phi_0} 
    \Bigl( \Psi(t) - \Phi_\text{diff} \Bigr) \biggr] \notag \\
    =  \sum_{m=1}^M \xi_m \ddot{\chi}_m(t),
\end{align}
where $C_\Psi \equiv C_\text{J} - K''/2$ and $L_\Psi \equiv K^{-1}$.
To find the auxiliary-mode equations of motion, we rewrite Eq.~\eqref{eq:chi_kernel} as
\begin{align}
    -\omega^2 \tilde{\chi}_m(\omega) &= -\xi_m \biggl[ 1 
    + \frac{\Omega_m^2}{\omega^2 - \Omega_m^2} \biggr] 
    \omega^2 \tilde{\Psi}(\omega),
\end{align}
which in the time domain, becomes
\begin{align} \label{eq:chi_EoM}
    \ddot{\chi}_m(t) &= \xi_m \ddot{\Psi}(t) - \Omega_m^2 \chi_m(t).
\end{align}
The Lagrangian that reproduces Eqs.~\eqref{eq:JJ_EoM_aux} and~\eqref{eq:chi_EoM} is
\begin{align} \label{eq:aux_mode_L}
    L_\text{S} &= \frac{1}{2} C_\Psi \dot{\Psi}^2 
    - \frac{\Psi^2}{2L_\Psi} 
    + E_\text{J} \cos\biggl[ \frac{2\pi}{\Phi_0} 
    \Bigl( \Psi - \Phi_\text{diff} \Bigr) \biggr] \notag \\
    &+ \sum_{m=1}^M \frac{1}{2} \Bigl[ \dot{\chi}_m^2 
    - \Omega_m^2 \chi_m^2 
    - 2\xi_m \dot{\chi}_m \dot{\Psi} \Bigr].
\end{align}
The resulting Lagrangian is equivalent to that introduced in Ref.~\cite{tuohino_multimode_2024}. There are, however, two key differences in the representation. Firstly, the interaction between the nonlinear mode and the auxiliary modes has been recast as a capacitive coupling. Secondly, the cutoff-dependent inductive renormalization terms are absent, which eliminates the strong dependence of the nonlinear mode transition frequency on the auxiliary-mode cutoff $M$, thereby enabling more efficient numerical calculations.

\subsection{Capacitively coupled drive}

The drive contribution is given in Eq.~\eqref{eq:L_D}. The flux field at the drive point, obtained from Eq.~\eqref{eq:aux_mode_flux_field}, takes the time-domain form
\begin{align} \label{eq:aux_mode_flux_field_time_t}
    \phi_{x}(x_\text{d},t) &= -F_{x}(x_\text{d},0)\,\Psi(t) 
    + \frac{1}{2}\partial_\omega^2 F_{x}(x_\text{d},\omega)
    \big|_{\omega=0} \ddot{\Psi}(t) \notag \\
    &\quad + \sum_{m=1}^M \frac{2f_{x,m}(x_\text{d})}{\Omega_m^3} 
    \frac{1}{\xi_m} \ddot{\chi}_m(t) \notag \\
    &= -F_{x}(x_\text{d},0)\,\Psi(t) 
    + \frac{1}{2}\partial_\omega^2 F_{x}(x_\text{d},\omega)
    \big|_{\omega=0} \ddot{\Psi}(t) \notag \\
    &\quad + \sum_{m=1}^M \frac{2f_{x,m}(x_\text{d})}{\Omega_m^3} 
    \biggl[ \ddot{\Psi}(t) 
    - \frac{\Omega_m^2}{\xi_m}\chi_m(t) \biggr] \notag \\
    &= -F_{x}(x_\text{d},0)\,\Psi(t) 
    - \sum_{m=1}^M \frac{2f_{x,m}(x_\text{d})}{\Omega_m\xi_m}
    \chi_m(t) \notag \\
    &\quad + \underbrace{\biggl[ 
    \frac{1}{2}\partial_\omega^2 F_{x}(x_\text{d},\omega)
    \big|_{\omega=0} 
    + \sum_{m=1}^M \frac{2f_{x,m}(x_\text{d})}{\Omega_m^3} 
    \biggr]}_{\to\, 0 \ \text{as} \ M \to \infty}
    \ddot{\Psi}(t) \notag \\
    &\approx -\Upsilon_x(x_\text{d})\,\Psi(t) 
    - \sum_{m=1}^M \zeta_{x,m}(x_\text{d})\,\chi_m(t),
\end{align}
where we have used Eq.~\eqref{eq:chi_EoM} and defined
\begin{align}
    \Upsilon_\alpha(u) &\equiv F_{\alpha}(u,0), \quad 
    \zeta_{\alpha,m}(u) \equiv \frac{2f_{\alpha,m}(u)}{\Omega_m\xi_m},
\end{align}
for $\alpha \in \{ x,y \}$. The vanishing of the $\ddot{\Psi}(t)$ term in the large-$M$ limit follows from the Mittag-Leffler theorem. The drive Lagrangian can thus be written as
\begin{align}
    L_\text{d} &= \frac{C_\text{d}}{2} \Bigl[ 
    \Upsilon_x(x_\text{d})\,\dot{\Psi}(t) 
    + \sum_{m=1}^M \zeta_{x,m}(x_\text{d})\,\dot{\chi}_m(t) 
    + V_\text{d}(t) \Bigr]^2.
\end{align}
Since the capacitive coupling to the drive line is weak, the renormalization terms may be neglected, giving
\begin{align}
    L_\text{d} &\approx C_\text{d} \Bigl[ 
    \Upsilon_x(x_\text{d})\,\dot{\Psi}(t) 
    + \sum_{m=1}^M \zeta_{x,m}(x_\text{d})\,\dot{\chi}_m(t) 
    \Bigr] V_\text{d}(t).
\end{align}

\subsection{Interaction with noisy magnetic flux}

Like the unimon, the multiunimon has a gradiometric loop structure that protects it from spatially large-scale magnetic-field fluctuations affecting both loops symmetrically. Fluctuations concentrated in a small area, however, generate a flux gradient between the two loops, which couples to the circuit through the flux field. We model this by adding a noisy component to the inductive energy density in the CPW Lagrangian of Eq.~\eqref{eq:L_CPW}, replacing
$$
\partial_u\phi_\alpha(u,t) \rightarrow \partial_u\phi_\alpha(u,t) - s \delta B_\text{diff}(u,t),
$$
where $\delta B_\text{diff}(x,t)$ is the difference between the magnetic flux densities of the two loops at position $u$, and $s$ is the dimension of the loop perpendicular to the central conductor.
The Lagrangian of this interaction is
\begin{align}
    L_{\Phi_{\text{diff}}} &= - \frac{s}{L_\ell}\sum_{\alpha \in \{ x,y \} }\int_{-l_\alpha}^{l_\alpha} \mathrm{d}u \ \delta B_\text{diff}(u,t)\, \partial_u \phi_\alpha(u,t) \notag \\
    &= \frac{s}{L_\ell} \delta B_\text{diff}(x_\text{J},t) \Psi(t) \notag \\
    &\quad + \frac{s}{L_\ell}\sum_{\alpha \in \{ x,y \} }\int_{-l_\alpha}^{l_\alpha} \mathrm{d}u \ \partial_u \delta B_\text{diff}(u,t)\, \phi_\alpha(u,t),
\end{align}
where we have used integration by parts, together with the grounding at the endpoints $\pm l_\alpha$ and the discontinuity of $\phi_\alpha$ at $x_\text{J}$. The second term depends on the spatial gradient of $\delta B_\text{diff}$ and is typically negligible. Assuming $\partial_u \delta B_\text{diff} (u,t)=0$, we obtain
\begin{align}
    L_{\Phi_{\text{diff}}} &\approx \frac{s}{L_\ell}\,
    \delta B_\text{diff}(x_\text{J},t)\,\Psi(t),
\end{align}
meaning the magnetic flux noise couples only to the nonlinear mode.

\subsection{Interactions with noisy capacitors}

We model dielectric losses by connecting a noisy voltage source $v_\alpha(u,t)$ in series with the CPW capacitance at each position $u$, leading to an interaction term
\begin{align}
    L_\text{diel} 
    &= - C_\ell \sum_{\alpha \in \{ x,y \} } \int_{-l_\alpha}^{l_\alpha} \mathrm{d}u\, v_\alpha(u,t) \, \dot{\phi}_\alpha(u,t) \notag \\
    &\approx C_\ell \sum_{\alpha \in \{ x,y \} } \int_{-l_\alpha}^{l_\alpha} \mathrm{d}u\, 
    v_\alpha(u,t) \notag \\
    & \qquad   \times \Biggl[ \Upsilon_\alpha(u)\,\dot{\Psi}(t) 
    + \sum_{m=1}^M \zeta_{\alpha,m}(u)\,\dot{\chi}_m(t) \Biggr],
\end{align}
where we have used a version of Eq.~\eqref{eq:aux_mode_flux_field_time_t} for general position $u$.

\subsection{Derivation of the Hamiltonian}

The classical Lagrangian of the full auxiliary-mode model including drive and 
noise sources is
\begin{align}\label{eq:lagrangian_parts}
    L &= L_\text{S} + L_\text{d} + L_\text{diel} + L_{\Phi_\text{diff}}.
\end{align}
This can be expressed compactly in matrix form as
\begin{align}\label{eq:lagrangian_matrix}
    L &= T - U,
\end{align}
where
\begin{align}\label{eq:T_U_defs}
    T &= \frac{1}{2}\dot{\mathbf{x}}^{\mathsf{T}} \mathbf{C}\,
    \dot{\mathbf{x}} 
    + \dot{\mathbf{x}}^{\mathsf{T}}(\mathbf{a} + \mathbf{b}), \\
    U &= \frac{1}{2}\mathbf{x}^{\mathsf{T}} \mathbf{L}^{-1}\mathbf{x} 
    - E_\text{J} \cos\biggl[ \frac{2\pi}{\Phi_0} 
    \Bigl( \Psi(t) - \Phi_\text{diff} \Bigr) \biggr] \notag \\ 
    &- \frac{s}{L_\ell}\,\delta B_\text{diff}(x_\text{J},t)\,\Psi(t)
\end{align}
are the kinetic and potential energy terms, respectively. The capacitance and inverse inductance matrices are
\begin{align}
    \mathbf{C} &=
    \begin{bmatrix}
        C_{\Psi}  & -\xi_{1}  & \cdots & -\xi_{M} \\
        -\xi_{1}  & 1         & \cdots & 0 \\
        \vdots    & \vdots    & \ddots & \vdots \\
        -\xi_{M}  & 0         & \cdots & 1
    \end{bmatrix}, \\
    \mathbf{L}^{-1} &=
    \begin{bmatrix}
        L^{-1}_{\Psi} & 0          & \cdots & 0 \\
        0             & \Omega_1^2 & \cdots & 0 \\
        \vdots        & \vdots     & \ddots & \vdots \\
        0             & 0          & \cdots & \Omega_M^2
    \end{bmatrix},
\end{align}
and the coordinate and coupling vectors are
\begin{align} \label{eq:gamma}
    \mathbf{x} =
    \begin{bmatrix}
        \Psi(t) \\ \chi_1(t) \\ \vdots \\ \chi_M(t)
    \end{bmatrix},
    \quad
    \bm{\gamma}_{\alpha}(u) =
    \begin{bmatrix}
        \Upsilon_\alpha(u) \\ \zeta_{\alpha,1}(u) \\ 
        \vdots \\ \zeta_{\alpha,M}(u)
    \end{bmatrix},
\end{align}
\begin{align}
    \mathbf{a} &= C_\text{d}\,V_\text{d}(t)\,\bm{\gamma}_{x}(x_\text{d}), 
    \label{eq:a_vec} \\
    \mathbf{b} &= C_\ell \sum_{\alpha \in \{ x,y \} }\int_{-l_\alpha}^{l_\alpha} \mathrm{d}u \,
    v_\alpha(u,t)\,\bm{\gamma}_{\alpha}(u). 
    \label{eq:b_vec}
\end{align}
The matrices $\mathbf{C}$ and $\mathbf{L}^{-1}$ are not genuine capacitance and inverse-inductance matrices, but quantities that serve a similar role. This is because the auxiliary-mode coordinates $\chi_m$ are not expressed in conventional flux units, but in a flux-like form scaled by a mode capacitance that the auxiliary-mode formalism does not fix to a unique value.

The canonical momenta $\mathbf{p} = \partial L / \partial \dot{\mathbf{x}}$ 
give
\begin{align}
    \mathbf{p} &= \mathbf{C}\,\dot{\mathbf{x}} + (\mathbf{a} + \mathbf{b}),
\end{align}
which can be inverted to give $\dot{\mathbf{x}} = \mathbf{C}^{-1}
(\mathbf{p} - \mathbf{a} - \mathbf{b})$. Applying the Legendre 
transformation $H = \dot{\mathbf{x}}^{\mathsf{T}}\mathbf{p} - L$ 
yields
\begin{align}\label{eq:Hamiltonian_matrix}
    H &= \frac{1}{2}(\mathbf{p} - \mathbf{a} - \mathbf{b})^{\mathsf{T}} 
    \mathbf{C}^{-1} (\mathbf{p} - \mathbf{a} - \mathbf{b}) + U.
\end{align}
The inverse capacitance matrix takes the form
\begin{align} \label{eq:inv_C}
    \mathbf{C}^{-1} &= \frac{1}{C_\text{eff}}
    \begin{bmatrix}
        1        & \xi_{1}       & \xi_{2}       & \cdots & \xi_{M} \\
        \xi_{1}  & C_\text{eff} + \xi_1^2 & \xi_1\xi_2 & \cdots & \xi_1\xi_M \\
        \xi_{2}  & \xi_2\xi_1   & C_\text{eff} + \xi_2^2 & \cdots & \xi_2\xi_M \\
        \vdots   & \vdots       & \vdots        & \ddots & \vdots \\
        \xi_{M}  & \xi_M\xi_1   & \xi_M\xi_2    & \cdots & C_\text{eff} + \xi_M^2
    \end{bmatrix},
\end{align}
where $C_\text{eff} = C_\Psi - \sum_{m=1}^M \xi_m^2$, and the 
canonical momentum vector is $\mathbf{p}^{\mathsf{T}} = \bigl[ Q,\; 
\Xi_1,\; \ldots,\; \Xi_M \bigr]$. Analogously to 
Eq.~\eqref{eq:lagrangian_parts}, we decompose the Hamiltonian as
\begin{align}\label{eq:Hamiltonian_parts}
    H &= H_\text{S} + H_\text{d} + H_\text{diel} 
    + H_{\Phi_\text{diff}},
\end{align}
where
\begin{align}
    H_\text{S} &= \frac{1}{2}\mathbf{p}^{\mathsf{T}} \mathbf{C}^{-1} 
    \mathbf{p} + \frac{1}{2}\mathbf{x}^{\mathsf{T}} \mathbf{L}^{-1}
    \mathbf{x} \notag \\
    &- E_\text{J} \cos\biggl[ \frac{2\pi}{\Phi_0} 
    \Bigl( \Psi - \Phi_\text{diff} \Bigr) \biggr], \\
    H_\text{d} &= -\mathbf{p}^{\mathsf{T}} \mathbf{C}^{-1} \mathbf{a}, \\
    H_\text{diel} &= -\mathbf{p}^{\mathsf{T}} \mathbf{C}^{-1} \mathbf{b}, \\
    H_{\Phi_\text{diff}} &= -\frac{s}{L_\ell}\,
    \delta B_\text{diff}(x_\text{J},t)\,\Psi(t).
\end{align}

\subsection{Rotation of the auxiliary-mode basis}

The charge gauge underlying Eq.~\eqref{eq:aux_mode_L} couples each auxiliary mode capacitively to the nonlinear mode. The Legendre transformation turns this into mutual couplings among the auxiliary modes, which appear as the off-diagonal entries of the inverse capacitance matrix in Eq.~\eqref{eq:inv_C}. In order to apply the iterative diagonalization scheme (see Appendix~\ref{app:NRG}), we perform a basis rotation on the auxiliary modes that removes these couplings and restores the star topology shown in Fig.~\hyperref[fig:circuit_schematic]{\ref*{fig:circuit_schematic}(b)}, in which the nonlinear mode acts as a central hub connected to the auxiliary modes as peripheral oscillators.

To this end, we apply transformations that diagonalize the 
auxiliary-mode basis. The new basis is defined as
\begin{align}\label{eq:aux_basis_new}
    \bm{\tilde{\chi}} &= \bm{\tilde{\Omega}}^{-1/2} \bm{O}^{\mathsf{T}} 
    \bm{L}_\text{aux}^{-1/2} \bm{\chi}, \\
    \bm{\tilde{\Xi}} &= \bm{\tilde{\Omega}}^{1/2} \bm{O}^{\mathsf{T}} 
    \bm{L}_\text{aux}^{1/2} \bm{\Xi},
\end{align}
where $\bm{O}$ is an orthogonal transformation satisfying
\begin{align}\label{eq:O_def}
    \bm{O}^{\mathsf{T}} \Bigl( \bm{L}_\text{aux}^{-1/2} 
    \bm{C}_\text{aux}^{-1} \bm{L}_\text{aux}^{-1/2} \Bigr) \bm{O} 
    = \bm{\tilde{\Omega}}^2 
    = \mathrm{diag}(\omega_1^2,\, \ldots,\, \omega_M^2).
\end{align}

Throughout this subsection, quantities carrying the subscript `aux' refer exclusively to the auxiliary harmonic modes, while the subscript $0$ denotes the nonlinear-mode degree of freedom, and vectors and matrices are partitioned accordingly.

The inverse relations are
\begin{align}\label{eq:aux_basis_new_inv}
    \bm{\chi} &= \bm{L}_\text{aux}^{1/2} \bm{O}\, 
    \bm{\tilde{\Omega}}^{1/2} \bm{\tilde{\chi}}, \\
    \bm{\Xi} &= \bm{L}_\text{aux}^{-1/2} \bm{O}\, 
    \bm{\tilde{\Omega}}^{-1/2} \bm{\tilde{\Xi}}.
\end{align}
In this basis, the classical system Hamiltonian takes the form
\begin{align} \label{eq:H_S_diag}
    H_\text{S} &= \frac{Q^2}{2C_\text{eff}} + \frac{\Psi^2}{2L_\Psi} 
    - E_\text{J} \cos\biggl[ \frac{2\pi}{\Phi_0} 
    \Bigl( \Psi - \Phi_\text{diff} \Bigr) \biggr] \notag \\
    &+ \sum_{m=1}^M \frac{\omega_m}{2} 
    \Bigl( \tilde{\Xi}_m^2 + \tilde{\chi}_m^2 \Bigr) 
    + \frac{Q}{C_\text{eff}} \sum_{m=1}^M 
    \tilde{\xi}_m\, \tilde{\Xi}_m,
\end{align}
where the coupling coefficients are defined as
\begin{align}
    \bm{\tilde{\xi}} = \bm{\tilde{\Omega}}^{-1/2} \bm{O}^{\mathsf{T}} 
    \bm{L}_\text{aux}^{-1/2} \bm{\xi}.
\end{align}
The drive Hamiltonian takes the form
\begin{align} \label{eq:H_D_diag}
    H_\text{d} &= -Q\, \tilde{a}_0 
    - \bm{\tilde{\Xi}}^{\mathsf{T}} \bm{\tilde{a}}_\text{aux},
\end{align}
where
\begin{align}
    \tilde{a}_0 &= \frac{ a_0 
    + \bm{\xi}^{\mathsf{T}} \bm{a}_\text{aux}}{C_\text{eff}}, 
    \label{eq:a_0_tilde} \\
    \bm{\tilde{a}}_\text{aux} &= \bm{\tilde{\Omega}}^{-1/2} 
    \bm{O}^{\mathsf{T}} \bm{L}_\text{aux}^{-1/2} 
    \left( \frac{a_0}{C_\text{eff}}\, \bm{\xi} + \bm{C}_\text{aux}^{-1} 
    \bm{a}_\text{aux} \right). \label{eq:a_aux_tilde}
\end{align}
Replacing $\bm{a}$ with $\bm{b}$, the dielectric 
Hamiltonian takes the analogous form
\begin{align} \label{eq:H_diel_diag}
    H_\text{diel} &= -Q\, \tilde{b}_0 
    - \bm{\tilde{\Xi}}^{\mathsf{T}} \bm{\tilde{b}}_\text{aux} \notag \\
    &= -C_\ell \sum_{\alpha \in \{ x,y \} }\int_{-l_\alpha}^{l_\alpha} \mathrm{d}u \, v_\alpha(u,t) \notag \\ 
    & \qquad \times \! \left[ Q\, \tilde{\gamma}_{\alpha,0}(u) 
    + \bm{\tilde{\Xi}}^{\mathsf{T}} \bm{\tilde{\gamma}}_{\alpha,\text{aux}}(u) 
    \right],
\end{align}
where we have defined
\begin{align}
    \tilde{\gamma}_{\alpha,0}(u) &= \frac{\gamma_{\alpha,0}(u) 
    + \bm{\xi}^{\mathsf{T}} \bm{\gamma}_{\alpha,\text{aux}}(u)}{C_\text{eff}}, 
    \label{eq:gamma_0_tilde} \\
    \bm{\tilde{\gamma}}_{\alpha,\text{aux}} (u) &= \bm{\tilde{\Omega}}^{-1/2} 
    \bm{O}^{\mathsf{T}} \bm{L}_\text{aux}^{-1/2} 
    \! \left[ \! \frac{\gamma_{\alpha,0}(u)}{C_\text{eff}}\, \bm{\xi} + \bm{C}_\text{aux}^{-1} 
    \bm{\gamma}_{\alpha,\text{aux}} (u) \! \right] \! . \label{eq:gamma_aux_tilde}
\end{align}

%\textcolor{blue}{TODO: Write down the definitions $H_\text{diel}$ more carefully.}

\subsection{Quantization of the Hamiltonian}\label{app:quantization}

The classical Hamiltonian in Eq.~\eqref{eq:H_S_diag} consists of $M$ harmonic oscillators capacitively coupled to a single nonlinear mode. We quantize the classical degrees of freedom by promoting them to operators, $\{Q, \Psi; \Xi_m, \chi_m\} \rightarrow \{\hat{Q}, \hat{\Psi}; \hat{\Xi}_m, \hat{\chi}_m\}$, expressed in dimensionless form as
\begin{align}
    \hat{Q} &= 2e\,\hat{n}, \quad 
    \hat{\Psi} = \frac{\Phi_0}{2\pi}\hat{\varphi}, \\ 
    \hat{\Xi}_m &= \ii\sqrt{\frac{\hbar}{2}}
    \bigl(\hat{a}_m^\dagger - \hat{a}_m\bigr), \quad 
    \hat{\chi}_m = \sqrt{\frac{\hbar}{2}}
    \bigl(\hat{a}_m^\dagger + \hat{a}_m\bigr),
\end{align}
satisfying the commutation relations $[\hat{\varphi}, \hat{n}] = \ii$ and $[\hat{a}_i, \hat{a}_j^\dagger] = \delta_{ij}$, with all other operator pairs commuting. With these operators, and assuming the sweet spot $\Phi_\text{diff} = \Phi_0/2$, the system Hamiltonian takes the form
\begin{align} \label{eq:H_S_dimensionless}
    \hat{H}_\text{S} &= 4E_C \hat{n}^2 + \frac{1}{2}E_L\hat{\varphi}^2 
    + E_\text{J}\cos\hat{\varphi} \notag \\
    &+ \sum_{m=1}^M \hbar\omega_m \hat{a}_m^\dagger \hat{a}_m 
    - \ii \hat{n} \sum_{m=1}^M \hbar g_m 
    \bigl(\hat{a}_m - \hat{a}_m^\dagger \bigr),
\end{align}
where we have defined the charging and inductive energies
\begin{align}
    E_C \equiv \frac{e^2}{2C_\text{eff}}, \qquad 
    E_L \equiv \left(\frac{\Phi_0}{2\pi}\right)^2 \frac{1}{L_\Psi},
\end{align}
and the coupling strength
\begin{align}
    g_m = \frac{4}{e} E_C \frac{\tilde{\xi}_m}{\sqrt{2\hbar}}.
\end{align}

The drive and noise Hamiltonians take the forms
\begin{align} 
    \hat{H}_\text{d} &= -2e\,\tilde{a}_0\,\hat{n} 
    - \ii\sum_{m=1}^M \tilde{a}_m \sqrt{\frac{\hbar}{2}} 
    \bigl(\hat{a}_m^\dagger - \hat{a}_m\bigr), \notag \\
    &= 2e V_\text{d}(t) \left[\,\eta_0\,\hat{n} 
    + \ii\sum_{m=1}^M \eta_m \bigl(\hat{a}_m^\dagger - \hat{a}_m\bigr) \right]\!, \label{eq:H_D} 
    \\
    \hat{H}_\text{diel} &= -2e\,\tilde{b}_0\,\hat{n} 
    - \ii\sum_{m=1}^M \tilde{b}_m \sqrt{\frac{\hbar}{2}} 
    \bigl(\hat{a}_m^\dagger - \hat{a}_m\bigr) \label{eq:H_diel} \\
    &= -C_\ell \sum_{\alpha \in \{ x,y \} }\int_{-l_\alpha}^{l_\alpha} \mathrm{d}u \, v_\alpha(u,t) \notag \\
    & \quad \times \! \biggl[ 2e\,\tilde{\gamma}_{\alpha,0}(u)\,\hat{n} \! + \! \ii\sum_{m=1}^M \tilde{\gamma}_{\alpha,m}(u) 
    \sqrt{\frac{\hbar}{2}} 
    \bigl(\hat{a}_m^\dagger - \hat{a}_m\bigr) \biggr], \notag \\
    \hat{H}_{\Phi_\text{diff}} &= -\frac{\Phi_0}{2\pi} 
    \frac{s}{L_\ell}\,\delta B_\text{diff}(x_\text{J},t)\,
    \hat{\varphi}, \label{eq:H_Phi_diff}
\end{align}
where $\tilde{a}_m$ and $\tilde{b}_m$ denote the $m$-th components of the rotated source vectors $\bm{\tilde{a}}_\text{aux}$ and $\bm{\tilde{b}}_\text{aux}$ defined in the previous subsection, and for the drive part, we have further defined $\eta_0 \equiv -\tilde{a}_0/V_\text{d}(t)$ and $\eta_m \equiv -\frac{\tilde{a}_m}{2eV_\text{d}(t)}\sqrt{\frac{\hbar}{2}}$. These coefficients are time-independent since $\tilde{a}_0$ and $\tilde{a}_m$ are both proportional to $V_\text{d}(t)$ through Eq.~\eqref{eq:a_vec}.

\subsection{Renormalization due to high-frequency modes}

Unimon-type circuits contain, in theory, an infinite number of modes due to their CPW structure, which poses a problem for numerical treatment typically addressed with a mode cutoff. This is an acceptable approach in most cases when the coupling is weak or the mode frequencies are strongly detuned. However, even weakly interacting modes that are not actively participating in the system dynamics---i.e. modes in which only the vacuum state is occupied---have a renormalizing effect on the low-energy subspace of the system. Accounting for these renormalizations is of great importance when the simulation objectives are sensitive to small energy shifts in the relevant eigenstates.

To account for these renormalizations, we transform into a displaced-oscillator basis via a unitary charge-shift of the form
\begin{align}\label{eq:displacement_U}
    \hat{U} = \exp \Biggl[ - \ii \, \hat{n} \! \sum_{m=M_\ell+1}^M \!
    \frac{g_m}{\omega_m} \bigl(\hat{a}_m^\dagger + \hat{a}_m\bigr) \Biggr],
\end{align}
where $M_\ell$ is a secondary cutoff marking the upper index of modes excluded from the displacement transformation. The transformed Hamiltonian, $\hat H_\text{S} \rightarrow \hat U^\dagger \hat H_\text{S} \hat U$, takes the form
\begin{align} \label{eq:H_S_displaced}
    \hat{H}_\text{S} &= 4\tilde{E}_C\,\hat{n}^2 
    + \frac{1}{2}E_L\hat{\varphi}^2 
    + E_\text{J}\cos\hat{\varphi}\cos\hat{\delta} \notag \\
    &- E_\text{J}\sin\hat{\varphi}\sin\hat{\delta} 
    + E_L\hat{\varphi}\,\hat{\delta} 
    + \frac{1}{2}E_L\hat{\delta}^2 \notag \\
    &+ \sum_{m=1}^{M} \hbar\omega_m\hat{a}_m^\dagger\hat{a}_m 
    - \ii \hat{n}\sum_{m=1}^{M_\ell} \hbar g_m 
    \bigl(\hat{a}_m - \hat{a}_m^\dagger \bigr),
\end{align}
where we have defined
\begin{align}
    \tilde{E}_C &= E_C - \! \! \sum_{m=M_\ell+1}^M 
    \! \! \frac{\hbar|g_m|^2}{4\omega_m}, \\
    \hat{\delta} &= \sum_{m=M_\ell+1}^M 
    \! \frac{g_m}{\omega_m} 
    \bigl(\hat{a}_m^\dagger + \hat{a}_m\bigr).
\end{align}
To approximate the effect of the high-frequency modes, we truncate their Hilbert space to the vacuum and single-excitation manifolds, the effect of which is captured by a Schrieffer--Wolff (SW) transformation. Since these modes have high frequencies, only their vacuum states are meaningfully occupied, with the single-excitation manifold contributing only through perturbative corrections.

The Hilbert space of the high-frequency modes consists of states
\begin{align}
    \ket{\bm{0}} \equiv \bigotimes_{m} \ket{0_m}, \quad
    \ket{\bm{1}_k} \equiv \ket{1_k} \bigotimes_{m\neq k} \ket{0_m},
\end{align}
where $m, k \in \{M_\ell + 1,\, \ldots,\, M\}$. Before evaluating the relevant matrix elements, we note that
\begin{align}
    \cos\hat{\delta} &= \bigotimes_m 
    \frac{\hat{D}_m\!\left(\ii\frac{g_m}{\omega_m}\right) 
    + \hat{D}_m\!\left(-\ii\frac{g_m}{\omega_m}\right)}{2}, \\
    \sin\hat{\delta} &= \bigotimes_m 
    \frac{\hat{D}_m\!\left(\ii\frac{g_m}{\omega_m}\right) 
    - \hat{D}_m\!\left(-\ii\frac{g_m}{\omega_m}\right)}{2\ii},\\
    \hat{D}_m(\beta_m) &= \exp\bigl( 
    \beta_m\hat{a}_m^\dagger - \beta_m^*\hat{a}_m \bigr),
\end{align}
where $\hat{D}_m$ is the displacement operator for mode $m$. Using the properties
\begin{align}
    \matrixel{0_m}{\hat{D}_m(\beta_m)}{0_m} 
    &= \ee^{-|\beta_m|^2/2}, \\
    \hat{D}_m^\dagger(\beta_m)\,\hat{a}_m^\dagger\,
    \hat{D}_m(\beta_m) 
    &= \hat{a}_m^\dagger + \beta_m^*,
\end{align}
we obtain
\begin{align}
    \matrixel{\bm{0}}{\cos\hat{\delta}}{\bm{0}} 
    &= \exp\biggl(-\frac{1}{2}\sum_m 
    \Bigl|\frac{g_m}{\omega_m}\Bigr|^2\biggr), \\
    \matrixel{\bm{0}}{\sin\hat{\delta}}{\bm{1}_k} 
    &= \frac{g_k}{\omega_k}
    \exp\biggl(-\frac{1}{2}\sum_m 
    \Bigl|\frac{g_m}{\omega_m}\Bigr|^2\biggr), \\
    \matrixel{\bm{0}}{\hat{\delta}}{\bm{1}_k} 
    &= \frac{g_k}{\omega_k},
\end{align}
with $\matrixel{\bm{0}}{\sin\hat{\delta}}{\bm{0}} = 
\matrixel{\bm{0}}{\cos\hat{\delta}}{\bm{1}_k} = 0$.
Applying the SW transformation to second order to eliminate the single-excitation manifold, and using the above matrix elements, the Hamiltonian projected onto the vacuum manifold of the high-frequency modes takes the form
\begin{align} \label{eq:H_S_renormalized}
    \hat{H}_\text{S} &= 4\tilde{E}_C\,\hat{n}^2 
    + \frac{1}{2}E_L\hat{\varphi}^2 
    + \tilde{E}_\text{J}\cos\hat{\varphi} \notag \\
    &- \sum_{k=M_\ell+1}^M \frac{|g_k|^2}{\hbar \omega_k^3} 
    \Bigl(E_L\hat{\varphi} 
    - \tilde{E}_\text{J}\sin\hat{\varphi}\Bigr)^2 \notag \\
    &+ \sum_{m=1}^{M_\ell} \hbar\omega_m\hat{a}_m^\dagger\hat{a}_m 
    - \ii \hat{n}\sum_{m=1}^{M_\ell} \hbar g_m 
    \bigl(\hat{a}_m - \hat{a}_m^\dagger \bigr),
\end{align}
where
\begin{align} \label{eq:EJ_renormalized}
    \tilde{E}_\text{J} \equiv E_\text{J}
    \exp\biggl(-\frac{1}{2}\sum_k 
    \Bigl|\frac{g_k}{\omega_k}\Bigr|^2\biggr)
\end{align}
is the Josephson energy renormalized by the zero-point fluctuations of the high-frequency modes.

The high-frequency mode displacement operator in Eq.~\eqref{eq:displacement_U} must also be applied to the drive and dielectric noise Hamiltonians, as was done for the system Hamiltonian above, and the result projected onto the vacuum state of the eliminated modes ($m > M_\ell$). Using the relations
\begin{align}\label{eq:a_shift}
    [\hat{n}, \hat{U}] &= 0, \quad
    \hat{U}^\dagger\,\hat{a}_m\,\hat{U} 
    = \hat{a}_m - \ii\frac{g_m}{\omega_m}\,\hat{n},
\end{align}
one finds that the eliminated high-frequency oscillators renormalize the coupling of the nonlinear mode to the external drive and dielectric noise sources, with corrections proportional to $g_m/\omega_m$.

%\textcolor{blue}{TODO: discuss how the pole expansion up to M modes already captures the dispersive frequency shift from modes $m > M$, manifesting as a renormalized charging energy $E_C$.
%}

\section{Iterative diagonalization} \label{app:NRG}

We present a numerical scheme for diagonalizing the multiunimon Hamiltonian over the $M_\ell$ auxiliary modes retained after the high-frequency renormalization in Eq.~\eqref{eq:H_S_renormalized}. By design, a multiunimon encodes at least three qubits across multiple coupled modes. Even coarse models retaining only a few modes are already computationally demanding, and such reductions are too crude for design optimization, even if the remaining non-participating modes are included perturbatively. In the basis used here, quantitatively reliable spectra typically require around $M_\ell \gtrsim 150$, making standard exact-diagonalization approaches impractical. We therefore adopt a truncated-basis iterative diagonalization method in the spirit of the numerical renormalization group~\cite{bulla_numerical_2005, bulla_numerical_2008}. In this mapping, the nonlinear mode serves as the impurity, or central hub, while the auxiliary harmonic modes act as the bath, incorporated one by one with controlled truncation at each step.

\subsection{Diagonalization of the system Hamiltonian}

\begin{figure*}[t]
\includegraphics[scale=1.18]{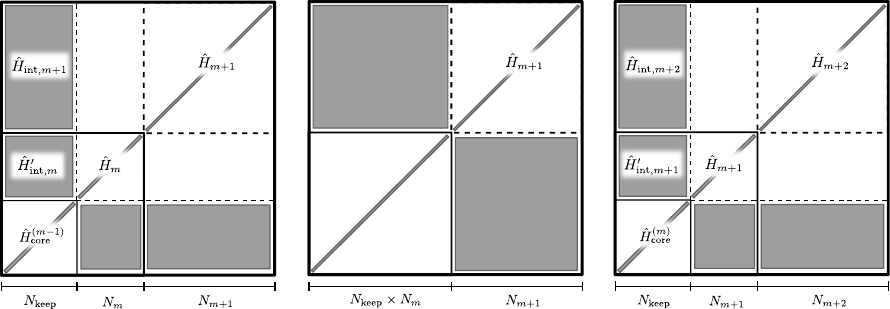}
\caption{\label{fig:iterative_diag}
Visualization of a single step in the iterative diagonalization scheme. Each auxiliary mode contributes a diagonal bare-mode block $\hat H_m$ and an off-diagonal coupling to the core; the prime on $\hat H'_{\text{int},m}$ marks the coupling expressed with the charge operator $\hat n$ in the truncated eigenbasis of the core. Left: the truncated core $\hat H_\text{core}^{(m-1)}$ (dimension $N_\text{keep}$), its rotated coupling $\hat H'_{\text{int},m}$ to mode $m$ (dimension $N_m$), and the next mode $\hat H_{m+1}$ (dimension $N_{m+1}$), whose coupling $\hat H_{\text{int},m+1}$ is still in the original basis. Middle: the $\hat H_\text{core}^{(m-1)} + \hat H_m$ block with the dimension $N_\text{keep}\times N_m$ is diagonalized, and the coupling to $\hat H_{m+1}$ is rotated accordingly. Right: truncating to $N_\text{keep}$ and relabeling the merged block as the new core $\hat H_\text{core}^{(m)}$ restores the structure of the left panel, and the next iteration begins.
}
\end{figure*}

Our iterative method begins by partitioning the system Hamiltonian into a nonlinear core $\hat{H}_0$ and a collection of mode-specific blocks $\hat{H}_m$. Each auxiliary mode is grouped together with its coupling to the nonlinear degree of freedom, while the intrinsic nonlinear term is retained in the core. The decomposition of Eq.~\eqref{eq:H_S_renormalized} reads
\begin{align} \label{eq:H_decomposition}
    \hat{H}_\text{S} &= \hat{H}_{0} + \sum_{m=1}^{M_\ell} \hat{H}_m, \\
    \hat{H}_{0} &= 4\tilde{E}_C\,\hat{n}^2 
    + \frac{1}{2}E_L\hat{\varphi}^2 
    + \tilde{E}_\text{J}\cos\hat{\varphi} \notag \\
    &\quad - \sum_{k=M_\ell+1}^M \frac{|g_k|^2}{\hbar\omega_k^3} 
    \Bigl(E_L\hat{\varphi} 
    - \tilde{E}_\text{J}\sin\hat{\varphi}\Bigr)^2, \\
    \hat{H}_m &= \hbar\omega_m\hat{a}_m^\dagger\hat{a}_m 
    - \ii \hat{n} \hbar g_m \bigl(\hat{a}_m - \hat{a}_m^\dagger \bigr).
\end{align}
Within each block, the first term is the diagonal bare-mode part, while the second is the off-diagonal interaction $\hat H_{\text{int},m} \equiv -\ii\hbar g_m\hat n(\hat a_m - \hat a_m^\dagger)$ that couples the mode to the core, as shown in Fig.~\ref{fig:iterative_diag}.

%The auxiliary modes are ordered by an estimate of their effective interaction strength — specifically the second-order scale $g_m^2/\omega_m^3$ — so that $m = 1$ is the most weakly coupled mode and $m = M_\ell$ the most strongly coupled.
We order the auxiliary modes by frequency, so that $m = 1$ is the highest-frequency mode and $m = M_\ell$ the lowest. We then diagonalize $\hat{H}_0$ to obtain eigenpairs $\{E_i^{(0)}, \ket{i^{(0)}}\}$ and retain a truncated set of low-energy eigenstates sufficient to meet a chosen accuracy target. In this eigenbasis, the charge operator is expanded as
\begin{align} \label{eq:n_expansion}
    \hat{n} &= \sum_{i,j} n_{ij}^{(0)}\,
    \ket{i^{(0)}}\!\bra{j^{(0)}}, \quad
    n_{ij}^{(0)} = \matrixelement{i^{(0)}}{\hat{n}}{j^{(0)}},
\end{align}
which is substituted into each coupling term,
\begin{align}
    \hat{n}\,(\hat{a}_m^\dagger - \hat{a}_m)
    = \sum_{i,j} n_{ij}^{(0)}\,
    \ket{i^{(0)}}\!\bra{j^{(0)}}\,
    (\hat{a}_m^\dagger - \hat{a}_m).
\end{align}
We choose a Fock-space cutoff $N_m$ for each auxiliary mode (typically $N_m > 8$ suffices) and form the first core Hamiltonian
\begin{align}
    \hat{H}_\text{core}^{(1)} = \hat{H}_{0} + \hat{H}_{1}.
\end{align}
Diagonalizing this yields eigenpairs $\{E_i^{(1)}, \ket{i^{(1)}}\}$, from which the lowest $N_\text{keep}^{(1)}$ eigenstates are retained. The charge operator is updated to the new eigenbasis,
\begin{align}
    \hat{n} = \sum_{i,j} n^{(1)}_{ij}\,
    \ket{i^{(1)}}\!\bra{j^{(1)}}, \quad
    n^{(1)}_{ij} = \matrixelement{i^{(1)}}{\hat{n}}{j^{(1)}},
\end{align}
and this representation is carried forward into all remaining interaction terms. In Fig.~\ref{fig:iterative_diag}, the coupling block evaluated with these updated matrix elements is the rotated block $\hat H'_{\text{int},m}$, whereas couplings to modes not yet incorporated remain in the original Fock basis. The procedure is then iterated: for $m = 2, \ldots, M_\ell$ we define
\begin{align}
    \hat{H}_\text{core}^{(m)} = 
    \hat{H}_\text{core}^{(m-1)} + \hat{H}_m,
\end{align}
diagonalize to obtain $\{E_i^{(m)}, \ket{i^{(m)}}\}$, truncate to 
$N_\text{keep}^{(m)}$ states, and update $\hat{n}$ accordingly. At step $m$ the diagonalized matrix has dimension $N_\text{keep}^{(m-1)} \times N_m$, illustrated by the merged block in the middle panel of Fig.~\ref{fig:iterative_diag}. After the final step,
\begin{align}
    \hat{H}_\text{core}^{(M_\ell)} = 
    \hat{H}_\text{core}^{(M_\ell - 1)} + \hat{H}_{M_\ell},
\end{align}
the diagonalization of $\hat{H}_\text{core}^{(M_\ell)}$ yields an approximate low-energy subspace of the full multimode Hamiltonian $\hat{H}_\text{S}$. Convergence is verified by increasing $\{N_m\}$ and $\{N_\text{keep}^{(m)}\}$ until the low-energy spectrum and relevant observables stabilize. The iterative process is visualized in Fig.~\ref{fig:iterative_diag}.

The auxiliary modes are included in order of descending frequency. This ordering may seem counterintuitive, but it improves convergence, for reasons rooted in the star topology of the Hamiltonian. Because the nonlinear mode couples to every auxiliary mode, its states carry the accumulated dressing from all modes included so far. Including the highest-frequency modes first therefore captures their dressing at a stage where the core basis is still largely made of nonlinear-mode excitations and little has been discarded. As lower-frequency modes are added at each step, states with explicit high-frequency excitations fall below the truncation threshold, but the dressing they produced is already encoded in the retained nonlinear-mode states and is passed on to the lower-frequency modes through the central coupling. In ascending order the situation reverses. The high-frequency modes enter last, when the core has already been truncated to a small low-energy subspace, and too few states with nonlinear-mode excitations remain to represent their dressing accurately.

%\textcolor{blue}{A figure visualizing the iterative diagonalization process should be added here.}

\subsection{Basis for environment coupling}

In our model, the multiunimon couples to a drive line and to dissipation channels arising from dielectric loss and external flux noise. As shown in Eqs.~\eqref{eq:H_D}, \eqref{eq:H_diel}, and~\eqref{eq:H_Phi_diff}, the environment acts on the system through the operators $\hat{O} \in \{\hat{n},\, \hat{\varphi};\, \hat{a}_m,\, \hat{a}_m^\dagger\}$. It is therefore convenient to represent each $\hat{O}$ in the eigenbasis of the full system Hamiltonian $\hat{H}_\text{S}$, which is approximated as  $\hat{H}_\text{core}^{(M_\ell)}$. Denoting the truncated eigenstates obtained after the final diagonalization by $\{\ket{i^{(M_\ell)}}\}$, we use the spectral expansion
\begin{align}
    \hat{O} &= \sum_{i,j} O^{(M_\ell)}_{ij}\,
    \ket{i^{(M_\ell)}}\!\bra{j^{(M_\ell)}}, 
    \label{eq:O_expansion} \\
    O^{(M_\ell)}_{ij} &= 
    \matrixelement{i^{(M_\ell)}}{\hat{O}}{j^{(M_\ell)}},
    \label{eq:O_matrix_elements}
\end{align}
where $i, j$ run over the retained low-energy subspace. These matrix elements serve as inputs for the drive term and the dissipative coupling terms that determine the Lindblad rates in the master-equation treatment.

\subsection{Efficiency of the method}

\begin{figure*}[t]
\includegraphics[scale=1.0]{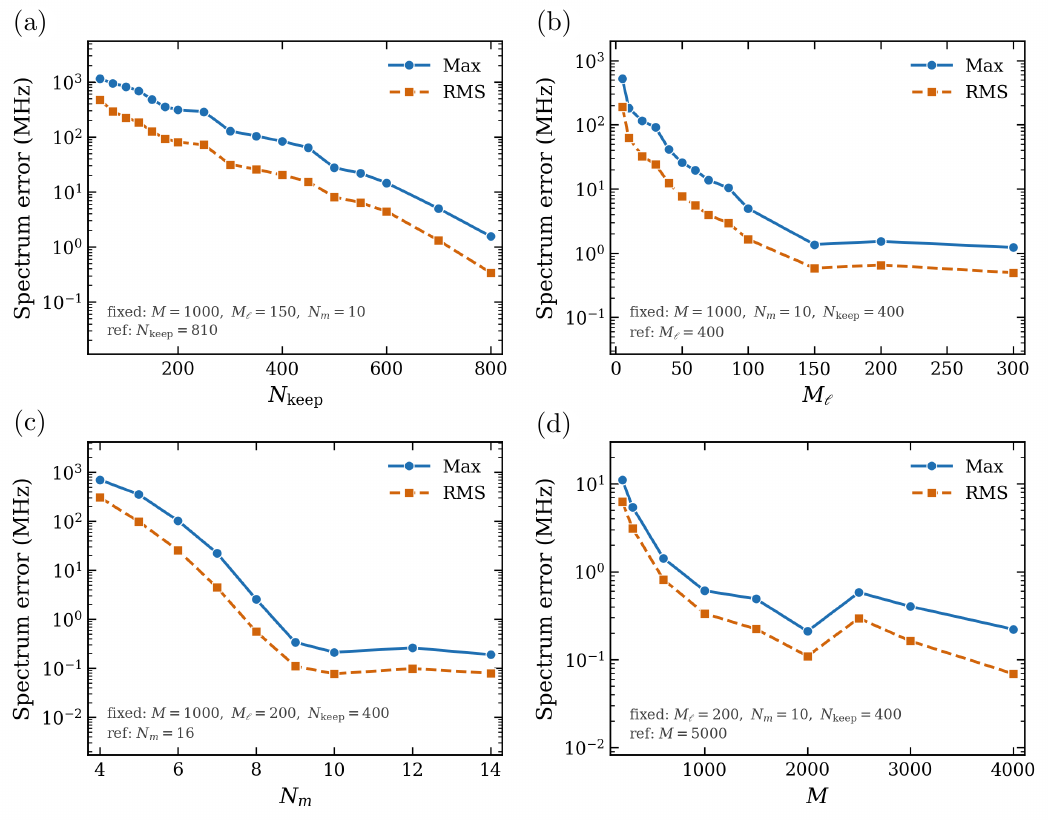}
\caption{\label{fig:convergence}
Convergence of the energy spectrum with respect to the relevant parameters: (a) $N_\text{keep}$, (b) $M_\ell$, (c) $N_m$, and (d) $M$. Each error is computed over the 50 lowest eigenstates, relative to a reference spectrum obtained at the large parameter value indicated in each panel. The blue curve shows the maximum deviation $\max\abs{E_i - E_{\text{ref},i}}$, and the orange curve the root-mean-square deviation $\sqrt{\sum_{i=0}^{49}\abs{E_i - E_{\text{ref},i}}^2/50}$, both over the same 50 states. Panels (a)–(c) test the truncation parameters of the iterative diagonalization, while (d) tests how strongly the low-energy spectrum still depends on the total mode number $M$ after the high-frequency renormalization of Appendix~\ref{app:Derivation}. The reference value of the swept parameter and the fixed parameters are given in each panel.
}
\end{figure*}

Balancing accuracy against computational cost requires careful control of the truncation levels $N_m$ and $N_\text{keep}^{(m)}$. At iteration $m$ the working Hilbert space has dimension $D^{(m)} = N_\text{keep}^{(m-1)}\, N_m$, so the Hamiltonian to be diagonalized is a $D^{(m)} \times D^{(m)}$ matrix. In typical runs we use $500 \leq N_\text{keep}^{(m-1)} \leq 800$ and $N_m \approx 10$, yielding matrix sizes up to $D^{(m)} \approx 8 \times 10^3$. Although these truncations can be chosen adaptively, increasing $N_m$ for more strongly coupled modes and enlarging $N_\text{keep}^{(m)}$ when necessary, we used uniform cutoff values for all $m \in \{ 1, \dots, M_\ell \}$. Figure~\ref{fig:convergence} indicates that the spectrum error for the 50-state low-energy subspace falls to~$\sim$~\SI{1}{\mega\hertz} at $N_\text{keep} = 800$, with the error plateauing near $M_\ell = 150$ and $N_m = 10$ and settling for $M \gtrsim 1000$. Convergence is faster still for the computational states at the lower end of the spectrum, which are the ones most relevant to gate operation.

We further accelerate the diagonalization by offloading the computationally intensive routines to a graphics processing unit (GPU). For the moderate matrix sizes encountered here, GPU parallelism is effective even on a laptop-class device. Using CuPy---a CUDA-backed Python array library---we achieve substantial speedups that enable more efficient exploration of the design parameter space during optimization.

\section{Decoherence channels}\label{app:Decoherence}

The multiunimon is susceptible to two primary noise sources: dielectric losses and external flux noise. The total relaxation time $T_{1,i}$ of each eigenstate, used in the gate simulations, arises from the sum of the transition rates to all other states,
\begin{equation}
    T_{1,i}^{-1} = \sum_{j \neq i} \left( \Gamma_{\text{diel,}ij} + \Gamma_{\text{flux},\,ij} + \Gamma^{1/f}_{\text{flux},\,ij} \right),
\end{equation}
where each $\Gamma_{ij}$ is the rate from state $i$ to state $j$, downward if $E_j < E_i$ and upward if $E_j > E_i$. The individual rates are estimated from Fermi's golden rule and derived in the following subsections.

\subsection{Dielectric loss}

We model dielectric loss by attaching, at each position $u$ along both CPW lines, a noisy voltage source $v_\alpha(u,t)$ in series with the local capacitance density $C_\ell$, with $\alpha \in \{ x,y \}$ labeling the line. Using Fermi's golden rule, the downward ($\downarrow$) and upward ($\uparrow$) transition rates between system eigenstates $\ket{i}$ and $\ket{j}$ with transition frequency~$\omega_{ij} = (E_i - E_j)/\hbar$ are
\begin{align}
\label{eq:Gamma_diel}
    \Gamma_{\text{diel,}ij}^{\downarrow/\uparrow}
    = \frac{1}{\hbar^2} \sum_{\alpha \in \{ x,y \}} \iint \mathrm{d}u\,& \mathrm{d}u'\;
    G_{\alpha,ji}(u) G_{\alpha,ij}(u') \notag \\
    & \times S^{\downarrow/\uparrow}_{vv}(u,u';\omega_{ij}),
\end{align}
where $G_{\alpha,ji}(u) \equiv \matrixelement{j}{\hat{G}_\alpha(u)}{i}$ are the matrix elements of the coupling operator
\begin{align}
    \hat{G}_\alpha(u) = -C_\ell\biggl[\hat{Q}\,\tilde{\gamma}_{\alpha,0}(u) + \sum_{m=1}^{M_\ell}\hat{\tilde{\Xi}}_m\,
    \tilde{\gamma}_{\alpha,m}(u)\biggr],
\end{align}
with $\tilde{\gamma}_{\alpha,\mu}(u)$ the position-dependent coupling profiles for channel $\mu = 0, 1, \ldots, M_\ell$ on line $\alpha$. We assume a bath that is spatially local, stationary, and uncorrelated between the two lines. This leads to an unsymmetrized voltage noise spectral density that factorizes as
\begin{align}
\label{eq:S_diel_spatial}
    S^{\downarrow/\uparrow}_{vv}(u,u';\omega)
    &= S^{\downarrow/\uparrow}_{vv}(\omega)\,\delta(u-u'), \\
\label{eq:S_diel_freq}
    S^{\downarrow/\uparrow}_{vv}(\omega)
    &= \frac{\hbar}{C_\ell\,Q_\text{diel}}
    \left[\coth\!\left(\frac{\hbar\omega}{2k_\text{B}T_{vv}}
    \right) \pm 1\right],
\end{align}
where the $+$ ($-$) sign corresponds to downward (upward) transitions, $Q_\text{diel}$ is the CPW dielectric quality factor, and $T_{vv}$ is the effective bath temperature for dielectric losses. Collecting the system matrix elements into a vector
\begin{equation}
\label{eq:p_vec}
    \mathbf{p}_{ij}^\dagger \equiv
    \bigl(Q_{ij},\;
    \tilde{\Xi}_{1,ij},\;\ldots,\;
    \tilde{\Xi}_{M_\ell,ij}\bigr),
\end{equation}
and the coupling profiles into
\begin{equation}
\label{eq:gamma_vec}
    \bm{\gamma}_\alpha(u) \equiv
    \bigl[\tilde{\gamma}_{\alpha,0}(u),\;
    \tilde{\gamma}_{\alpha,1}(u),\;\ldots,\;
    \tilde{\gamma}_{\alpha,M_\ell}(u)
    \bigr]^{\mathsf{T}},
\end{equation}
the rate in Eq.~\eqref{eq:Gamma_diel} can be written in the compact quadratic form
\begin{align}
\label{eq:Gamma_diel_mat}
    \Gamma_{\text{diel,}ij}^{\downarrow/\uparrow}
    &= \frac{C_\ell^{\,2}}{\hbar^2}\;
    \mathbf{p}_{ij}^\dagger\;
    \bm{\mathcal{K}}^{\downarrow/\uparrow}(\omega_{ij})\;
    \mathbf{p}_{ij}, \\
\label{eq:K_diel_def}
    \bm{\mathcal{K}}^{\downarrow/\uparrow}(\omega)
    &\equiv \! \! \!  \sum_{\alpha \in \{ x,y \}} \! \iint \mathrm{d}u\, \mathrm{d}u' \bm{\gamma}_\alpha(u) \bm{\gamma}_\alpha^\dagger(u') S^{\downarrow/\uparrow}_{vv}(u,u';\omega).
\end{align}
Under the locality assumption in Eq.~\eqref{eq:S_diel_spatial}, the kernel reduces to
\begin{align}
\label{eq:K_diel_local}
    \bm{\mathcal{K}}^{\downarrow/\uparrow}(\omega) &= S^{\downarrow/\uparrow}_{vv}(\omega)\;\bm{\Lambda}, \\
\label{eq:Lambda_overlap}
    \Lambda_{\mu\nu} &= \sum_{\alpha \in \{ x,y \}} \int \mathrm{d}u\; \tilde{\gamma}_{\alpha,\mu}^*(u)\,\tilde{\gamma}_{\alpha,\nu}(u),
\end{align}
where $\bm{\Lambda}$ is a Hermitian overlap matrix encoding the spatial structure of the coupling channels. Together, Eqs.~\eqref{eq:Gamma_diel_mat}--\eqref{eq:Lambda_overlap} determine all emission and absorption rates between system eigenstates. In our simulations we use $Q_\text{diel} = 3.5 \times 10^5$ and $T_{vv} =$~\SI{25}{\milli\kelvin}.

\subsection{Flux noise}

Flux noise arises from uncontrolled fluctuations of the magnetic flux threading the device loops. Owing to its gradiometric layout, the branched multiunimon suppresses common-mode flux fluctuations that shift both loops equally, but remains sensitive to differential fluctuations that modify the flux difference between the loops. Two physically distinct sources dominate: (i) current noise in the on-chip flux-bias line, whose spectrum is well described by an ohmic model, and (ii) environmental magnetic-field noise with a characteristic $1/f$ spectrum.

The coupling to differential flux noise derived in Appendix~\ref{app:Derivation} can be written in terms of the nonlinear mode as
\begin{align}
\label{eq:H_Phi_diff_main}
    \hat{H}_{\Phi_{\text{diff}}}(t) &= -\frac{2\pi}{\Phi_0}\, E_{\text{loop}}\,\delta\Phi_{\text{diff}}(t) \hat{\varphi},
\end{align}
where
\begin{align}
    E_{\text{loop}}
    &= \frac{(\Phi_0/2\pi)^2}{L_\ell(x_\text{b} + l_x + l_y)},
    \label{eq:E_loop} \\
\label{eq:Phi_diff_def}
    \delta\Phi_{\text{diff}}(t)
    &= A_{\text{loop}}\;\delta B_{\text{diff}}(x_\text{J},t),\\
\label{eq:A_loop_def}
    A_{\text{loop}}
    &= s\,(x_\text{b} + l_x + l_y).
\end{align}
Here $A_{\text{loop}}$ is an estimate of the effective loop area, $\delta B_{\text{diff}}(x_\text{J},t)$ is the differential magnetic-field fluctuation at the junction, $E_\text{loop}$ is the loop inductive energy in the distributed model, and $l_x$, $l_y$ are the half-lengths of the two CPW branches. Only the nonlinear degree of freedom $\hat{\varphi}$ is directly coupled to the flux noise. We note that junction--loop asymmetry ($y_\text{b} \neq 0$) breaks the gradiometric symmetry and increases the susceptibility to flux noise. However, for simplicity, we do not consider this in the relaxation-rate estimates presented here.
%We do not include this correction in the relaxation-rate estimates presented here, but it can be incorporated by updating the coupling prefactor in Eq.~\eqref{eq:H_Phi_diff_main}.

Using Fermi's golden rule, the downward and upward transition rates are
\begin{align}
\label{eq:Gamma_flux_FGR_general}
    \Gamma^{\downarrow/\uparrow}_{\text{flux},\,ij}
    &= \frac{1}{\hbar^2}\,
    \Bigl|\matrixelement{i}
    {\frac{\partial\hat{H}_{\Phi_{\text{diff}}}}
    {\partial\,\delta\Phi_{\text{diff}}}}{j}\Bigr|^{\!2}
    \;S^{\downarrow/\uparrow}_{\Phi\Phi}(\omega_{ij}).
\end{align}
Substituting Eq.~\eqref{eq:H_Phi_diff_main}, this takes the compact form
\begin{align}
\label{eq:Gamma_flux_compact}
    \Gamma^{\downarrow/\uparrow}_{\text{flux},\,ij}
    &= \left(\frac{2\pi\,E_{\text{loop}}}{\Phi_0\,\hbar}\right)^{\!2}
    |\varphi_{ij}|^2\;
    S^{\downarrow/\uparrow}_{\Phi\Phi}(\omega_{ij}),
\end{align}
where $S^{\downarrow/\uparrow}_{\Phi\Phi}(\omega)$ are the unsymmetrized flux-noise spectral densities.

\subsubsection{Ohmic flux noise}

For flux noise sourced by the flux-bias line, the noise spectrum is related to the current-noise spectrum via the mutual inductance $M_I$,
\begin{equation}
    S^{\downarrow/\uparrow}_{\Phi\Phi}(\omega)
    = M_I^2\,S^{\downarrow/\uparrow}_{II}(\omega)
    = \frac{\hbar\omega\,M_I^2}{R_I}
    \biggl[\coth\!\biggl(\frac{\hbar\omega}{2k_\text{B}T_{\Phi\Phi}}
    \biggr) \pm 1\biggr],
\end{equation}
where $R_I$ is the effective line resistance and the $+$ ($-$) sign corresponds to downward (upward) transitions. Substituting into Eq.~\eqref{eq:Gamma_flux_compact} gives
\begin{equation}
    \Gamma^{\downarrow/\uparrow}_{\text{flux},\,ij}
    = \frac{4\pi^2}{\Phi_0^2}\;
    \frac{E_{\text{loop}}^{2}M_I^2}{\hbar R_I}\;
    |\varphi_{ij}|^2\,\omega_{ij}
    \biggl[\coth\!\biggl(\frac{\hbar\omega_{ij}}{2k_\text{B}T_{\Phi\Phi}}
    \biggr) \pm 1\biggr],
\end{equation}
where we have used $M_I =$~\SI{1.3}{\pico\henry}, $R_I =$~\SI{26}{\ohm} and $T_{\Phi\Phi} =$~\SI{25}{\milli\kelvin}. We have assumed that the flux and dielectric baths share the same effective temperature, denoted by $T$ in Table~\ref{tab:design}.

\subsubsection{\texorpdfstring{$1/f$}{1/f} flux noise}

For environmental flux noise, the $1/f$ spectral density is symmetric ($S^\downarrow = S^\uparrow$) and takes the form
\begin{equation}
    S^{1/f}_{\Phi\Phi}(\omega)
    = \frac{2\pi\,A_{\Phi_\text{diff}}^{2}}{|\omega|},
\end{equation}
where $A_{\Phi_\text{diff}}$ is the noise amplitude specified at $1~\mathrm{Hz}$. Substituting into Eq.~\eqref{eq:Gamma_flux_compact} yields
\begin{equation}\label{eq:Gamma_1overf}
    \Gamma^{1/f}_{\text{flux},\,ij}
    = 8\pi^{3}\,\frac{E_{\text{loop}}^{2}}{\hbar^{2}}\,
    \frac{A_{\Phi_\text{diff}}^{2}}{\Phi_{0}^{2}}\,
    \frac{|\varphi_{ij}|^{2}}{|\omega_{ij}|},
\end{equation}
where we have set $A_{\Phi_\text{diff}} = \SI{15}{\micro{\Phi_0}}$. This is the relaxation contribution of the $1/f$ noise, which drives transitions at the qubit frequencies. 

\subsection{Pure dephasing due to \texorpdfstring{$1/f$}{1/f} noise}

Experiments on the unimon qubit indicate that pure dephasing from $1/f$ flux noise can be a significant contributor to decoherence~\cite{hyyppa_unimon_2022}. We generalize the quadratic dephasing model developed for the unimon at the sweet spot $\varphi_\text{diff} = \pi$~\cite{duda_parameter_2025,mukkula_optimizing_nodate} to the multiunimon. In a two-level subspace spanned by $\{\ket{i},\ket{j}\}$, the coherence envelope exhibits a characteristic non-exponential decay,
\begin{equation}
\label{eq:fij_def}
    %f_{ij}(t) = \frac{1}{\sqrt{1 - 2\ii\,\kappa_{ij} \bigl(2\pi A_{\Phi_\text{diff}}^{2}\bigr)\, t\,\ln\!\dfrac{1}{\omega_\text{ir}\,t}}},
    f_{ij}(t) = \frac{1}{\sqrt{1 - 2\ii\,\kappa_{ij}  A_{\Phi_\text{diff}}^{2} t\ln\!\dfrac{1}{\omega_\text{ir}\,t}}},
\end{equation}
where $\omega_\text{ir}$ is the infrared cutoff frequency and $\kappa_{ij}$ is the second derivative of the transition frequency $\omega_{ij} = (E_i - E_j)/\hbar$ with respect to the differential flux,
\begin{equation}
\label{eq:kappa_def}
    %\kappa_{ij} \equiv \frac{\partial^{2}\omega_{ij}}{\partial\Phi_\text{diff}^{2}} = \frac{8\pi^{2}}{\Phi_0^2}\, \frac{E_{\text{loop}}^{2}}{\hbar^{2}} \left(\sum_{k\neq i}\frac{|\varphi_{ik}|^{2}}{\omega_{ik}} - \sum_{l\neq j} \frac{|\varphi_{jl}|^{2}}{\omega_{jl}}\right).
    \kappa_{ij} \equiv \frac{\partial^{2}\omega_{ij}}{\partial\Phi_\text{diff}^{2}} = \frac{4\pi^{2}}{\Phi_0^2}\, \frac{E_{\text{loop}}^{2}}{\hbar^{2}} \left(\sum_{k\neq i}\frac{|\varphi_{ik}|^{2}}{\omega_{ik}} - \sum_{l\neq j} \frac{|\varphi_{jl}|^{2}}{\omega_{jl}}\right).
\end{equation}

Equation~\eqref{eq:fij_def} is valid in the regime $t < 2\bigl(\kappa_{ij}A_{\Phi_\text{diff}}^{2}\bigr)^{-1}$, which covers the time scales relevant for fast gate operations. Following Ref.~\cite{duda_parameter_2025}, we avoid the infrared divergence of the $1/f$ spectrum by setting $\omega_\text{ir} = 2\pi\times 1~\mathrm{kHz}$, and define the corresponding pure dephasing quasirate as
\begin{align}
    \tilde \Gamma_{\varphi, ij}^{1/f} \equiv 2 \kappa_{ij} A_{\Phi_\text{diff}}^2,
\end{align}
which is related to the power law decay given in Eq.~\eqref{eq:fij_def}.

Unlike the other noise sources, the pure dephasing due to $1/f$ flux noise is not included in the average-gate-infidelity simulations. However, it does not appear to significantly contribute to the fidelities. We characterize each pair by the dephasing time corresponding to its quasirate, $\tilde T_{\varphi,ij}^{1/f} = 1/\tilde \Gamma_{\varphi,ij}^{1/f}$. Across the computational subspace, the shortest is $\tilde T_{\varphi,01}^{1/f} =$ \SI{34.9}{\micro\second}, and the average over all pairs is \SI{231.3}{\micro\second}. Because the quasirate overestimates the true non-exponential decay of $f_{ij}(t)$, these times are worst-case estimates.

\section{Estimation of the gate fidelity} \label{app:fid_estimation}

We quantify gate performance using the average gate fidelity (AGF), a standard metric for comparing an ideal target unitary process $\hat U_0$ to a noisy physical implementation described by a quantum channel $\mathcal G(\rho)=\sum_k \hat G_k \rho \hat G_k^\dagger$. Projecting onto the computational subspace with $\hat P$, the AGF can be expressed as~\cite{pedersen_fidelity_2007}
\begin{align}
\label{eq:F_avg_def}
& \bar{\mathcal F} =\frac{\mathrm{Tr}\!\bigl(\sum_k \hat M_k^\dagger \hat M_k\bigr) + \sum_k \bigl|\mathrm{Tr}(\hat M_k)\bigr|^2}{d_P(d_P+1)}, \\
&\hat M_k \equiv \hat U_0^\dagger \hat P \hat G_k \hat P,
\end{align}
where $d_P=\mathrm{Tr}(\hat P)$ is the dimension of the computational subspace. For the efficient evaluation of the AGF in the full Hilbert space, we use the approach presented in Ref.~\cite{cabrera_average_2007}. However, the circuit parameter optimization loop requires an even more efficient approach.

\subsection{Two-level reduction for pairwise transitions}

Evaluating Eq.~\eqref{eq:F_avg_def} in the full Hilbert space is computationally expensive, with the main bottleneck being the computation of the noisy quantum channel $\mathcal{G}$. We therefore estimate the AGF by reducing the problem to a set of independent two-level sectors formed by pairs of eigenstates. Specifically, we let
\begin{align}
\mathcal{C} &= \bigl\{\{\ket{p},\ket{q}\}:\; \ket{p},\ket{q}\in\mathcal{H}_\text{comp}\bigr\}, \notag
\end{align}
denote all pairs within the computational subspace and 
\begin{align}
\mathcal{L} &= \bigl\{\{\ket{j},\ket{l}\}:\; \ket{j}\in\mathcal{H}_\text{comp},\; \ket{l}\in\mathcal{H}_\text{leak}\bigr\}, \notag
\end{align}
denote all pairs formed by one computational and one leakage state. We then compute a two-level average gate fidelity for each pair in $\mathcal{C}\cup\mathcal{L}$ and combine these to form an estimate of $\bar{\mathcal{F}}$. This is a drastic simplification of the general case, in which the pairs cannot be treated independently. For ranking device designs by expected AGF, however, the approach is computationally efficient while remaining sufficiently accurate, making it a suitable objective function for design optimization.

\subsection{Average gate fidelities with the pairwise model}

For a typical multiunimon device, the computational subspace contains eight basis states connected by native three-qubit controlled-controlled rotations ($\mathrm{CC}\mathcal{R}$). Consider a gate designed to implement a $\pi$ rotation between a specific pair $\{\ket{p},\ket{q}\}\in\mathcal{C}$. The corresponding two-level Kraus operators take the form
\begin{equation}
\label{eq:G_k}
    \hat{G}_k = \hat{D}_k\,\hat{U}_{t_g}^\dagger,
\end{equation}
where $\hat{U}_{t_g}$ captures the coherent drive dynamics and the operators $\{\hat{D}_k\}$ encode amplitude damping and dephasing of the pair. In the rotating frame of the drive, the time evolution takes the standard Rabi form
\begin{align}
\label{eq:rabi_evolution}
    \hat{U}_{t_g} \dot=
    \begin{pmatrix}
        c - \ii s_z & -\ii \ee^{-\ii\phi} s_x \\[2pt]
        -\ii \ee^{\ii\phi} s_x & c + \ii s_z
    \end{pmatrix},
\end{align}
where we have introduced the shorthand notation
\begin{align}
    c &\equiv \cos\!\frac{\Omega t_g}{2}, \qquad
    s_z \equiv \frac{\Delta}{\Omega}\sin\!\frac{\Omega t_g}{2}, \\
    s_x &\equiv \frac{W}{\Omega}\sin\!\frac{\Omega t_g}{2}, \qquad
    \Omega \equiv \sqrt{W^2+\Delta^2},
\end{align}
with $W$ the on-resonant Rabi rate, $\phi$ the drive phase, and $\Delta = \omega_d - \omega_{pq}$ the detuning. The symbol $\dot =$ denotes a matrix representation in a given basis, which in this case is $\{\ket{p},\ket{q}\}$.

We model the decoherence in this subspace with a set of Kraus operators combining generalized amplitude damping and pure dephasing,
\begin{align}
\label{eq:kraus_op1}
\hat D_{0,pq} &\dot= \sqrt{\frac{(1+\lambda_{pq}) (1-\bar{n}_{pq})}{2}}
\begin{pmatrix} 1 & 0\\[2pt] 0 & \sqrt{1-p_{pq}}\end{pmatrix}, \\
\hat D_{1,pq} &\dot= \sqrt{\frac{(1-\lambda_{pq}) (1-\bar{n}_{pq})}{2}}
\begin{pmatrix} 1 & 0\\[2pt] 0 & -\sqrt{1-p_{pq}}\end{pmatrix}, \\
\hat D_{2,pq} &\dot= \sqrt{\frac{(1+\lambda_{pq}) \bar{n}_{pq}}{2}}
\begin{pmatrix} \sqrt{1-p_{pq}} & 0\\[2pt] 0 & 1\end{pmatrix}, \\
\hat D_{3,pq} &\dot= \sqrt{\frac{(1-\lambda_{pq}) \bar{n}_{pq}}{2}}
\begin{pmatrix} \sqrt{1-p_{pq}} & 0\\[2pt] 0 & -1\end{pmatrix}, \\[4pt]
%\hat D_{4,pq} &\dot= \sqrt{(1-p_{pq})\bar{n}_{pq}}
\hat D_{4,pq} &\dot= \sqrt{(1-\bar{n}_{pq})p_{pq}}
\begin{pmatrix} 0 & 1\\[2pt] 0 & 0 \end{pmatrix}, \\
\label{eq:kraus_op2}
\hat D_{5,pq} &\dot= \sqrt{p_{pq}\bar{n}_{pq}}
\begin{pmatrix} 0 & 0\\[2pt] 1 & 0 \end{pmatrix},
\end{align}
where $p_{pq}\equiv 1 - \ee^{-\Gamma_{pq} t_g}$ accounts for population transfer between states $q$ and $p$, and $\lambda_{pq}\equiv \ee^{-\Gamma_{\phi,pq}t_g}$ accounts for pure dephasing~\cite{Nielsen_Chuang_2010}. The rates $\Gamma_{pq} \equiv \Gamma_{pq}^\uparrow + \Gamma_{pq}^\downarrow$ and $\Gamma_{\phi,pq}$ are constructed from the noise channels in Appendix~\ref{app:Decoherence}.

\paragraph{Desired transition.}
For the target $\pi$ pulse between $\ket{p}$ and $\ket{q}$ we set $t_g=\pi/W_{pq}$, $\Delta=0$, and $\phi=0$, for which the coherent evolution reduces to the ideal gate,
%which sets the coherent part of the quantum process to be equal to the ideal gate,
\begin{equation}
\label{eq:rabi_evolution_d}
\hat U^{(d)}_{t_g} \;=\; \hat U^{(d)}_0 \;\dot=\; -\,\ii
\begin{pmatrix} 0 & 1 \\[2pt] 1 & 0 \end{pmatrix}.
\end{equation}
Evaluating Eq.~\eqref{eq:F_avg_def} with the Kraus set in Eqs. \eqref{eq:kraus_op1}-\eqref{eq:kraus_op2} yields the average gate fidelity for the desired transition
\begin{equation}
\label{eq:F_avg_def_d}
\bar{\mathcal{F}}^{(d)}
=\frac{1}{6}\Bigl(
3 + \ee^{-t_g/T_{1,pq}} + 2 \ee^{-t_g/T_{2,pq}}
\Bigr),
\end{equation}
with
\begin{equation}
\label{eq:T1_T2}
T_{1,pq}^{-1} = \Gamma_{pq}^\uparrow + \Gamma_{pq}^\downarrow,\qquad
T_{2,pq}^{-1} = \frac{\Gamma_{pq}^\uparrow + \Gamma_{pq}^\downarrow}{2} + \Gamma_{\phi,pq}.
\end{equation}
In this idealized case, the only contribution to the infidelity is decoherence during the gate.

\paragraph{Undesired transitions within the computational subspace.}

The above-used two–level approach is applied to all pairs in $\mathcal C$ not targeted by the drive, which we call spectator pairs. Ideally these transitions remain idle during the drive, but in practice the drive can induce off-resonant transitions within the computational manifold and ac Stark phase shifts. Both are coherent errors, but the ac Stark phase shifts can be compensated by virtual phase gates, leaving off-resonant transitions as the dominant drive-induced error for spectator pairs.

To account for the ac Stark phase shifts, we take the ideal operation on the spectator pair to be a $Z$ rotation,
\begin{equation}
\label{eq:U_0_z_corr}
    \hat{U}_0^{(u)}(\theta)
    \dot= \begin{pmatrix} \ee^{-\ii\theta/2} & 0 \\[2pt] 
    0 & \ee^{\ii\theta/2} \end{pmatrix}.
\end{equation}
Inserting $\hat{U}_0^{(u)}(\theta)$ into Eq.~\eqref{eq:F_avg_def} yields
\begin{align} \label{eq:F_avg_u}
&\bar{\mathcal{F}}^{(u)} = \frac{3 + (1 - 2s_x^2)\,\ee^{-t_g/T_{1,pq}}}{6} \notag \\
&+ \frac{\ee^{-t_g/T_{2,pq}}\!\bigl[ (c^2-s_z^2)\cos(\Delta t_g - \theta) + 2cs_z\sin(\Delta t_g - \theta) \bigr]}{3},
\end{align}
where the uncorrected AGF corresponds to $\theta = 0$. The optimal phase correction is
\begin{equation}
\label{eq:theta_opt}
    \theta_\text{opt} = \Delta t_g 
    - 2\arctan\!\left(\frac{s_z}{c}\right),
\end{equation}
which maximizes Eq.~\eqref{eq:F_avg_u}, giving
\begin{align} \label{eq:F_avg_u_z_corr}
    \bar{\mathcal{F}}^{(u,\mathrm{Z\text{-}corr})} 
    &= \frac{3 + \ee^{-t_g/T_{1,pq}} 
    + 2\ee^{-t_g/T_{2,pq}}}{6} \notag \\
    &- \frac{2s_x^2\Bigl(\ee^{-t_g/T_{1,pq}} 
    + \ee^{-t_g/T_{2,pq}}\Bigr)}{6}.
\end{align}
The first line of Eq.~\eqref{eq:F_avg_u_z_corr} represents the baseline AGF in the absence of the off-resonant drive, while the second line gives the penalty proportional to $s_x^2$, corresponding to the off-resonant population transferred within the computational subspace.

\paragraph{Undesired transitions leaking outside the computational subspace.}

Finally, we consider pairs in $\mathcal{L}$ describing drive-induced leakage outside the computational subspace. The target unitary is the identity $\hat{U}_0^{(\ell)} = \hat{I}$, and following Eq.~\eqref{eq:F_avg_def}, we compare it against the two-level channel for each pair $\{\ket{j},\ket{l}\}$. Since only $\ket{j}$ belongs to $\mathcal{H}_\text{comp}$, the projector is $\hat{P} = \ketbra{j}{j}$, giving $\hat{M}_k = \matrixelement{j}{\hat{G}_k}{j}\ketbra{j}{j}$. Using this and $d_P = 1$, Eq.~\eqref{eq:F_avg_def} reduces to
\begin{equation}
    \bar{\mathcal{F}}^{(\ell)}
    = \sum_k \bigl|\matrixelement{j}{\hat{G}_k}{j}\bigr|^2.
\end{equation}
Inserting the Kraus operators yields two cases depending on whether the computational state $\ket{j}$ has lower or higher energy than the leakage state $\ket{l}$,
\begin{align}
\label{eq:F_avg_leak_jl_low}
    \bar{\mathcal{F}}^{(\ell,\, j < l)}
    &= 1 - \bar{n}_{jl}\bigl(1 - \ee^{-t_g/T_{1,jl}}\bigr) 
    - s_x^2\,\ee^{-t_g/T_{1,jl}}, \\ 
\label{eq:F_avg_leak_jl_high}
    \bar{\mathcal{F}}^{(\ell,\, j > l)}
    &= 1 - (1-\bar{n}_{jl})\bigl(1 - \ee^{-t_g/T_{1,jl}}\bigr) 
    - s_x^2\,\ee^{-t_g/T_{1,jl}},
\end{align}
where $\bar{n}_{jl}$ is the thermal equilibrium excited-state population of the two-level sector $\left\{ \ket{j}, \ket{l} \right\}$,
\begin{equation} \label{eq:n_bar}
    \bar{n}_{jl} \equiv 
    \frac{\Gamma_{jl}^\uparrow}{\Gamma_{jl}^\uparrow
    +\Gamma_{jl}^\downarrow} 
    = \frac{1}{\ee^{\hbar\omega_{jl}/(k_\text{B}T)} + 1}.
\end{equation}
The two cases arise because the fidelity metric tracks only population in $\ket{j}$, so incoherent transitions between $\ket{j}$ and $\ket{l}$ contribute asymmetrically. When $\ket{j}$ has higher energy than $\ket{l}$, population can spontaneously decay into the leakage state even at low temperatures, leading to enhanced incoherent loss. When the energy of $\ket{j}$ is lower, incoherent leakage requires thermal excitation and is therefore suppressed at low temperatures.

Pure dephasing does not appear in Eqs.~\eqref{eq:F_avg_leak_jl_low} and~\eqref{eq:F_avg_leak_jl_high}, for the same reason as above: the fidelity is computed solely over the one-dimensional subspace spanned by $\ket{j}$, making it insensitive to phase information.

We found that better estimates are often obtained by using an approximate form for
$s_x$ corresponding to a sine-squared pulse shape. After the optimal phase correction, the pairwise fidelities depend on the coherent evolution only through $s_x$, since $s_z$ and $c$ are eliminated by the optimal phase correction. The only change needed is therefore to replace $s_x$ with
\begin{equation} \label{eq:sx_cos}
    s_x^\text{sin} = \min{ \! \left( \frac{4 \pi^2 W}{\abs{\Delta} \abs{\Delta^2 t_g^2 - 4 \pi^2}}, \, 1 \right) }.
\end{equation}
This form is better aligned with the frequency spectrum of the pulse used in the gate simulations and yields more accurate estimates.

\subsection{Aggregating pairwise fidelities}

To combine the pairwise average fidelities into a single figure of merit for the full computational subspace, we convert each pairwise average fidelity $\bar{\mathcal{F}}^{(\mathcal X)}$ to an entanglement infidelity
\begin{equation}
\label{eq:entanglement_infidelity}
\mathcal E_e^{(\mathcal X)} \;=\; \frac{d_{\mathcal X}+1}{d_{\mathcal X}}\bigl(1-\bar{\mathcal{F}}^{(\mathcal X)}\bigr),
\end{equation}
where $d_{\mathcal X}$ is the dimension of the subspace on which the pairwise channel acts (here $d_{\mathcal C}=2$ for computational pairs and $d_{\mathcal L}=1$ for comp--leakage pairs), and $\mathcal X\in\{\mathcal C,\mathcal L\}$. Equation~\eqref{eq:entanglement_infidelity} follows from the standard relation $\bar{\mathcal{F}}=(d\,\mathcal F_e+1)/(d+1)$ with $\mathcal E_e=1-\mathcal F_e$.

Let $d \equiv \dim\mathcal H_\text{comp}$ denote the dimension of the full computational subspace, distinct from the per-pair dimension $d_{\mathcal X}$ above. Because the $\binom{d}{2} = d(d-1)/2$ computational pairs share states and are not independent, we normalize by their number to avoid over-counting, giving $w_{\mathcal C} = \frac{2}{d(d-1)}$. Likewise, for comp–leakage pairs we use $w_{\mathcal L}=1/d$, which corresponds to assigning equal total weight to the $d$ computational states, i.e., each computational state contributes its aggregate leakage. Treating the pairwise contributions as small and independent, a first-order additive estimate for the overall entanglement infidelity on the computational subspace is
\begin{equation}
\label{eq:E_e_full_comp}
\mathcal E_e^{(\mathrm{comp})} \;\approx\; \sum_{\{p,q\}\in\mathcal C} w_{\mathcal C}\,\mathcal E_{e,pq} \;+\; \sum_{\{j,l\}\in\mathcal L} w_{\mathcal L}\,\mathcal E_{e,jl}.
\end{equation}
Transforming back yields an approximation to the average gate fidelity on the full computational subspace:
\begin{equation}
\label{eq:F_avg_full_comp}
\bar{\mathcal{F}}^{(\mathrm{comp})} \;\approx\; \frac{d\bigl(1-\mathcal E_e^{(\mathrm{comp})}\bigr)+1}{d+1}.
\end{equation}

\subsection{Average gate-set infidelity}

In a typical multiunimon device, the computational manifold ($d=8$) admits twelve native three-qubit connections activated by the drive. Applying Eq.~\eqref{eq:E_e_full_comp} to each native gate $i=1,\dots,12$ produces $\mathcal E_{e,i}^{(\mathrm{comp})}$. To summarize the overall quality of the gate set with a single metric, we define the weighted average gate-set entanglement infidelity
\begin{equation}
\label{eq:AGSI}
\overline{\mathcal E}_\mathrm{set} \;=\; \frac{\sum_{i=1}^{12} w_i\,\mathcal E_{e,i}^{(\mathrm{comp})}} {\sum_{i=1}^{12} w_i},
\end{equation}
where $w_i\!\ge\!0$ are user-chosen importance weights. In this work we take uniform weights, $w_i=1$, so that $\overline{\mathcal E}_\mathrm{set}$ reduces to the simple arithmetic mean over the twelve native gates.

\section{Selection of the computational subspace} \label{app:cube_search}

The idea behind leakage-aware encoding is to use the pairwise infidelity metric of Appendix~\ref{app:fid_estimation} to select the computational states with the lowest estimated gate-set infidelity for given circuit parameters, forming the inner level of the optimization loop. Eight computational states define the vertices of a three-dimensional cube, where each of the twelve edges corresponds to a distinct native gate, and we refer to a candidate assignment of the states as \emph{the cube}. The number of low-energy eigenstates from which the states are selected typically ranges from 50 to 100, and hence an exhaustive testing of all candidate cubes is not viable. Combining a prefilter, a layered beam search, and an exact branch-and-bound stage yields a high-quality encoding at a tractable cost. Below, we give an overview of the search process.

Part of what makes this task challenging is that the cost assigned to each edge of the cube, or each transition, depends on the vertices, or states, not directly involved in the edge. We refer to this as the context dependence of the edge cost. Mathematically,
\begin{equation}
\label{eq:partial_cube}
c(i,j;\mathcal{S}) = \sum_{k \in \mathcal{S}} \varepsilon_{ijk},
\end{equation}
where $i$ and $j$ denote the computational states connected by the edge, $\mathcal S$ is the set of computational states in the cube, and $\varepsilon_{ijk}$ is the infidelity contribution of state $k$ to the gate driven on edge $(i,j)$. We form the per-state contributions from the pairwise entanglement infidelities of Appendix~\ref{app:fid_estimation}. Other definitions may be used as well. The edge costs are combined into a single metric representing the cost of a complete or partial cube. We use the sum of squares over the edge costs, Eq.~\eqref{eq:partial_cube}, which penalizes cubes containing a single poor edge more strongly than the mean and performs well in practice.

During the search, $\mathcal S$ is filled by the states starting from the fixed ground state such that the edge costs are updated each time a state is added. An edge that initially looks promising can suffer from leakage caused by a computational state added later. Provided the per-state contributions are non-negative, as holds for the infidelity-based definition used here, the edge cost is monotonically non-decreasing as states are added. The cost of a partial cube is therefore a lower bound on the cost of every complete cube containing it, and a partial cube whose cost already exceeds the best complete cube found can be discarded without examining its completions. This monotonicity is the key property that makes the search tractable. The main steps of the search are as follows.

\paragraph*{1. Fixed ground state.}
We fix the computational ground state $\ket{000}$ to the physical ground state of the system. The remaining seven states are determined by the search.

\paragraph*{2. Prefilter and feasibility constraints.}
The candidate pool is first reduced by pruning states that are unsuitable for any cube. Because the relaxation time of a state is independent of the encoding context, states with very short $T_1$ are removed outright. In addition, the microwave control electronics limit the addressable frequency range, and in this work, we require every gate transition to lie between \SI{1}{\giga\hertz} and \SI{15}{\giga\hertz}. This frequency constraint, together with an energy cutoff on the highest states involved, is enforced throughout the search rather than only at the prefilter stage.

\paragraph*{3. Layered beam search.}
The cube is assembled layer by layer, $G \rightarrow S \rightarrow D \rightarrow T$, where the layers group the states by their number of excitations, from zero for $G$ to three for $T$. States are placed one at a time. Throughout, the search maintains a beam of $W$ lowest-cost partial cubes. At each placement, every beam member is extended by all admissible candidate states. The edge costs are updated with the contribution of each new state, all extensions from all beam members are ranked together, and the $W$ lowest-cost partial cubes are retained. A greedy assembly ($W = 1$) is unreliable because of the context dependence of the edge costs: the best partial cube need not to extend to the best complete cube. The beam hedges against this by keeping alternative partial solutions alive, and we found $W = 100$ a good compromise between accuracy and efficiency. The layered beam search results in a feasible cube that performs better than most other candidates. However, because the ranking favors partial cubes that perform well at the early layers, the beam can miss the true optimum, which motivates the exact search of the following step.

\paragraph*{4. Branch and bound.}
Although the beam search does not guarantee optimal solution, its result serves as an initial upper bound for an exact search. The branch-and-bound stage explores the tree of partial cubes and discards a branch as soon as its accumulated cost meets the bound: by the monotonicity of the cost, no completion of such a branch can improve on the best cube found.

Similar to the layered beam search, it builds the cube layer by layer, starting from the $S$ layer. Every branch with a partial cost exceeding that of the best complete cube found so far is eliminated immediately. Because the layered beam search already supplies a strong initial candidate, most branches are discarded early, which keeps the search practical. If the search runs to completion, the returned cube is optimal over the prefiltered candidate set. Otherwise, it returns the best candidate found within the chosen time limit.

\begin{widetext}

\section{Kernels for the multiunimon} \label{app:kernel_defs}

% --- Optional short macros (helps readability) ---
\newcommand{\lx}{l_x}
\newcommand{\ly}{l_y}
\newcommand{\xB}{x_{\text{b}}}
\newcommand{\yB}{y_{\text{b}}}
\newcommand{\xJ}{x_{\text{J}}}
\newcommand{\Cl}{C_\ell}

The kernels of the multiunimon and other long definitions used in Appendix~\ref{app:Derivation} are defined as
\begin{align}
N_{x,1}(x,\omega) &= 
\Big(
   -2\cos\!\big[k_\omega (l_x + 2l_y - \xJ)\big]
   -\cos\!\big[k_\omega (l_x + 2l_y - 2\xB + \xJ)\big]
   +\cos\!\big[k_\omega (l_x - 2(l_y + \xB) + \xJ)\big] \notag \\
&\qquad\quad
   +\cos\!\big[k_\omega (l_x - \xJ - 2\yB)\big]
   +\cos\!\big[k_\omega (l_x - \xJ + 2\yB)\big]
\Big)\sin\!\big[k_\omega (l_x + x)\big], \\[6pt]
N_{x,2}(x,\omega) &= 
\cos\!\big[k_\omega (l_x + \xJ)\big]\Big(
  2\sin\!\big[k_\omega (l_x + 2l_y - x)\big]
 +\sin\!\big[k_\omega (l_x - 2l_y + x - 2\xB)\big] \notag \\
&\qquad\qquad
 -\sin\!\big[k_\omega (l_x + 2l_y + x - 2\xB)\big]
 +\sin\!\big[k_\omega (-l_x + x - 2\yB)\big]
 +\sin\!\big[k_\omega (-l_x + x + 2\yB)\big]
\Big), \\[6pt]
N_{x,3}(x,\omega) &= 
-4\cos\!\big[k_\omega (l_x + \xJ)\big]\,
\sin\!\big[k_\omega (-l_x + x)\big]\,
\sin\!\big[k_\omega (-l_y + \yB)\big]\,
\sin\!\big[k_\omega (l_y + \yB)\big],  \\[6pt]
N_{y,1}(y,\omega) &= 
-4\cos\!\big[k_\omega (l_x + \xJ)\big]\,
\sin\!\big[k_\omega (-l_x + \xB)\big]\,
\sin\!\big[k_\omega (l_y + y)\big]\,
\sin\!\big[k_\omega (-l_y + \yB)\big], \\[6pt]
N_{y,2}(y,\omega) &= 
-4\cos\!\big[k_\omega (l_x + \xJ)\big]\,
\sin\!\big[k_\omega (-l_x + \xB)\big]\,
\sin\!\big[k_\omega (-l_y + y)\big]\,
\sin\!\big[k_\omega (l_y + \yB)\big], \\[6pt]
D(\omega) &= 
2\sin\!\big[2k_\omega (l_x + l_y)\big]
 -\sin\!\big[2k_\omega (l_y - \xB)\big]
 -\sin\!\big[2k_\omega (l_y + \xB)\big]
 -\sin\!\big[2k_\omega (l_x - \yB)\big]
 -\sin\!\big[2k_\omega (l_x + \yB)\big],
\end{align}

\begin{align}
F_{x,1}(x,0) &= \frac{(l_x + x)\left[-l_y(2\,l_x + l_y - 2\,x_{\text{b}}) + y_{\text{b}}^2\right]}{2\,l_x\,l_y\,(l_x + l_y) - 2\,l_y\,x_{\text{b}}^2 - 2\,l_x\,y_{\text{b}}^2} = -\frac{(l_x + x)\left[-l_y(2\,l_x + l_y - 2\,x_{\text{b}}) + y_{\text{b}}^2\right]}{2 \mathcal D}, \\[1em]
F_{x,2}(x,0) &= \frac{\,l_y\!\left[-l_y\,x + 2(x - x_{\text{b}})\,x_{\text{b}} + l_x\,(l_y - 2x + 2x_{\text{b}})\right] + (-l_x + x)\,y_{\text{b}}^2}{2\,l_x\,l_y\,(l_x + l_y) - 2\,l_y\,x_{\text{b}}^2 - 2\,l_x\,y_{\text{b}}^2} \\
&= -\frac{\,l_y\!\left[-l_y\,x + 2(x - x_{\text{b}})\,x_{\text{b}} + l_x\,(l_y - 2x + 2x_{\text{b}})\right] + (-l_x + x)\,y_{\text{b}}^2}{2 \mathcal D}, \\[1em]
F_{x,3}(x,0) &= \frac{(l_x - x)\,(l_y - y_{\text{b}})\,(l_y + y_{\text{b}})}{2\,l_x\,l_y\,(l_x + l_y) - 2\,l_y\,x_{\text{b}}^2 - 2\,l_x\,y_{\text{b}}^2} = - \frac{(l_x - x)\,(l_y - y_{\text{b}})\,(l_y + y_{\text{b}})}{2 \mathcal D}, \\[1em]
F_{y,1}(y,0) &= \frac{(l_x - x_{\text{b}})\,(l_y + y)\,(l_y - y_{\text{b}})}{2\,l_x\,l_y\,(l_x + l_y) - 2\,l_y\,x_{\text{b}}^2 - 2\,l_x\,y_{\text{b}}^2} = -\frac{(l_x - x_{\text{b}})\,(l_y + y)\,(l_y - y_{\text{b}})}{2 \mathcal D}, \\[1em]
F_{y,2}(y,0) &= \frac{(l_x - x_{\text{b}})\,(l_y - y)\,(l_y + y_{\text{b}})}{2\,l_x\,l_y\,(l_x + l_y) - 2\,l_y\,x_{\text{b}}^2 - 2\,l_x\,y_{\text{b}}^2}
=-\frac{(l_x - x_{\text{b}})\,(l_y - y)\,(l_y + y_{\text{b}})}{2 \mathcal D},
\end{align}

\begin{align}
K = \frac{1}{2L_\ell}\frac{l_y\,(2\,l_x + l_y - 2\,x_{\text{b}}) - y_{\text{b}}^{2}}
{l_y\,[\,l_x\,(l_x + l_y) - x_{\text{b}}^{2}\,] - l_x\,y_{\text{b}}^{2}} = -\frac{1}{2L_\ell} \frac{L}{\mathcal D},
\end{align}

% --- Final expression (compact form) ---
\begin{align}
K''=\frac{\Cl\big(\mathcal D\,P + L\,Q\big)}{3\,\mathcal D^{2}},
\end{align}

% --- Define common factors ---
\[
\begin{aligned}
\mathcal D &\equiv -\,\lx\ly(\lx+\ly) + \ly \xB^2 + \lx \yB^2,\\[0.25em]
L &\equiv \ly(2\lx+\ly-2\xB) - \yB^2,\\[0.25em]
P &\equiv 4\lx^3\ly
      + 2\lx\ly\!\left[2\ly^2 + 3(\xB-\xJ)^2\right]
      + \ly\!\left[\ly^3 - 4\ly^2\xB - 4\xB^3 + 6\xB^2\xJ + 3(\ly-2\xB)\xJ^2\right] \\
  &\quad\; - 3\xJ^2 \yB^2 - \yB^4
      + 3\lx^2\!\left[\ly(\ly - 2\xB + 2\xJ) - \yB^2\right],\\[0.25em]
Q &\equiv \lx^4\ly + 2\lx^2\ly^3
      - \ly \xB^2\!\left(2\ly^2 + \xB^2\right)
      + 2\lx^3(\ly-\yB)(\ly+\yB)
      + \lx(\ly^4 - \yB^4).
\end{aligned}
\]

\end{widetext}

\bibliography{bibliography}% Produces the bibliography via BibTeX.

\end{document}